\documentclass[useAMS,usenatbib]{mnras}

\usepackage{graphics}
\usepackage{amssymb}
\usepackage{amsfonts}
\usepackage{wasysym}
\usepackage{epsfig}
\usepackage{breakurl}

\usepackage{longtable}
\usepackage{times}
\usepackage{mathptmx}
\usepackage[usenames,dvipsnames]{color}

\newcommand{\s}{\scriptsize}

\title[The magnetosphere of HR5907]
{A combined multiwavelength VLA/ALMA/{\em Chandra} study unveils the complex magnetosphere of the B-type star HR5907}
\author[P. Leto et al.]
{P. Leto$^{1}$ \thanks{E-mail: pleto@oact.inaf.it},
C. Trigilio$^{1}$,
L. M. Oskinova$^{2}$,
R. Ignace$^{3}$,
C. S. Buemi$^{1}$,
G. Umana$^{1}$,
\newauthor A. Ingallinera$^{1}$,
F. Leone$^{4,1}$,
N. M. Phillips$^{5,6}$,
C. Agliozzo$^{5}$,
H. Todt$^{2}$,
L. Cerrigone$^{6,7}$
\\
$^{1}$INAF - Osservatorio Astrofisico di Catania, Via S. Sofia 78, 95123 Catania, Italy\\
$^2$Institute for Physics and Astronomy, University Potsdam, 14476 Potsdam, Germany\\
$^3$Department of Physics \& Astronomy, East Tennessee State University, Johnson City, TN 37614, USA\\
$^4$Universit\'{a} degli studi di Catania, Via S.Sofia 78, I-95123 Catania, Italy\\
$^5$European Southern Observatory, Alonso de C\'{o}rdova 3107, Vitacura, Santiago, Chile\\
$^6$Joint ALMA Observatory, Alonso de C\'{o}rdova 3107, Vitacura, Santiago, Chile\\
$^7$Associated Universities, Inc., Av. Nueva Costanera 4091, Suite 502, Vitacura, Santiago, Chile\\
}
\begin{document}

\date{}

\pagerange{\pageref{firstpage}--\pageref{lastpage}} \pubyear{}

\maketitle

\label{firstpage}

\begin{abstract}
We present new radio/millimeter measurements of the hot magnetic star HR\,5907 
obtained with the VLA and ALMA interferometers. 
We find that HR\,5907 is the most radio luminous early type star in the cm-mm band {among those presently known.} 
Its multi-wavelength radio light curves are strongly variable 
with an amplitude that increases with radio frequency. 
The radio emission can be explained by the populations of the non-thermal electrons accelerated 
in the current sheets on the outer border of the magnetosphere of this fast rotating magnetic star. 
We classify HR\,5907 as another member of the growing class of strongly 
magnetic fast rotating hot stars where the gyro-synchrotron emission mechanism efficiently operates in their magnetospheres.
The new radio observations of HR\,5907 are combined with archival X-ray 
data to study the physical condition of its magnetosphere. 
The X-ray spectra of HR\,5907 show tentative evidence for the presence of non-thermal spectral component. 
We suggest that non-thermal X-rays originate {a stellar X-ray aurora} 
due to {streams of non-thermal electrons} impacting on the stellar surface. 
Taking advantage of the relation between the spectral indices of the X-ray power-law spectrum 
and the non-thermal electron energy distributions, we perform 3-D modeling of the radio emission for HR\,5907. 
The {wavelength-dependent} radio light-curves probe magnetospheric 
layers at different heights above the stellar surface. 
A detailed comparison between simulated and observed radio light-curves leads us 
to conclude that the stellar magnetic field of HR\,5907 is likely non-dipolar,
{providing further indirect} evidence of the complex magnetic field topology of HR\,5907. 
\end{abstract}


\section{Introduction}

At the top of the main sequence, magnetism has an incidence rate of
about $10\%$  \citep{grunhut_etal12b,fossati_etal15}.  
{The study of the magnetic fields of early-type stars} has received 
extraordinary attention in the form large legacy programs, like the
Magnetism in Massive Stars (MiMeS) project \citep{wade_etal16} and
the B-Fields in OB Stars (BOB) collaboration \citep{morel_etal14,morel_etal15}.  
The overall field topology of the hot magnetic stars is usually shaped like a dipole,
and in many cases, the polar field strength reaches the kiloGauss (kG) level.  
The existence of strong magnetic fields in hot stars
can significantly affect the circumstellar environment. 
In fact, the early-type main sequence stars are hot enough to give rise to a radiatively
driven stellar wind \citep*{lucy_solomon70,castor_etal75}.
A strong stellar magnetic field can force the ionized
wind component to flow along the magnetic field lines. 
In the magnetic equator, the field lines are closed and the wind
plasma can accumulate \citep{walborn74,shore_etal87,shore_brown90},
whereas at the higher magnetic latitudes the stellar wind freely
propagates \citep{shore87,leone93}.

The interactions between stellar winds and magnetospheres of early-type stars were described 
by the magnetically confined wind shock (MCWS) model \citep{babel_montmerle97}.  
In the framework of MCWS, the wind streams arising from
the two opposite stellar hemispheres collide close to the magnetic equatorial plane, 
giving rise to shocks that produce plasma heating and consequent  X-rays emission 
(see \citealp{ud-doula_naze16} and references therein).

If the stellar magnetic field is strong enough, 
the thermal plasma that accumulates at low magnetic latitudes
is forced to rigidly co-rotate with the star, {well-described} by the
Rigidly Rotating Magnetosphere (RRM) model \citep{townsend_owocky05}.
{Moreover, magnetohydrodynamics \citep{ud-doula_etal06,ud-doula_etal08}
and rigid-field hydrodynamics \citep{townsend_etal07}
simulations confirm the formation of a stable disk,
as predicted by the RRM model.}
The prototype of a hot magnetic star with a RRM is $\sigma$\,Ori\,E
\citep{groote_hunger97,townsend_etal05}.
{Several $\sigma$\,Ori\,E analogues since discovered are:
$\delta$\,Ori\,C \citep{leone_etal10};
HD\,176582 \citep{bohlender_monin11};
HR\,5907 \citep{grunhut_etal12a};
HR\,7355 \citep{rivinius_etal13};
HD\,23478 \citep{eikenberry_etal14,sikora_etal15};
HD\,345439 \citep{eikenberry_etal14,wisniewski_etal15,hubrig_etal17};
HD\,35502 \citep{sikora_etal16};
HD\,164492C \citep{wade_etal17}.
}

The plasma in an RRM is {mainly subjected} to two opposing forces:  gravitational and the centrifugal. 
The first drives the plasma to infall toward the star, whereas the latter is outward.  
Rotation supplies the centrifugal support to the co-rotating plasma.  
Stellar rotation is thus the key parameter to produce large centrifugal
magnetospheres (CM) \citep{maheswaran_cassinelli09,petit_etal13}.
{There are presumably other forces acting on the magnetospheric plasma
(e.g. radiative driving and the Lorentz force), but these are neglected in the RRM model.}
Far from the star, close to the Alfv\'{e}n radius, the magnetic
confinement ceases, and the trapped thermal plasma breaks the
magnetic field lines giving rise to current sheets where electrons
can be accelerated to relativistic energies.
This non-thermal electron population, 
moving toward the star, travels through magnetospheric regions
with increasing field strength and radiates a continuum
radio spectrum via the incoherent gyro-synchrotron emission mechanism
\citep*{drake_etal87,linsky_etal92,leone_etal94}.  
As a consequence, the radio emission from stable co-rotating magnetospheres 
varies as a function of the stellar rotation \citep{leone91,leone_umana93}.  
Multi-wavelength radio light curves for the total intensity and for the fraction of the
circularly polarized radio emission have been successfully simulated 
using a 3D model that, sampling the space surrounding the star,
numerically solves the radiative transfer equation for the
gyro-synchrotron emission \citep{trigilio_etal04,leto_etal06}.

{Such a model provides} a powerful tool
to investigate the physical conditions in stellar magnetospheres.
In particular, we conducted radio observations,
using the Karl G.\ Jansky Very Large Array (VLA),
of a representative sample of magnetic hot stars.
The sample stars are characterized by different physical properties,
such as the geometry of their magnetospheres, 
the strength of their magnetic fields, and the stellar rotation speed.
The first object analyzed from the radio point-of-view
was HR\,7355 \citep{leto_etal17a}.
In that case the synergistic radio and X-ray diagnostic techniques
led to a scenario in which all the data could be understood within
a single model realization.
In particular the X-ray spectrum of HR\,7355
{showed indication of} a non-thermal component that was explained
as the signature of the non-thermal electron population
that impacted with the stellar surface, close to the polar caps, to 
produce X-ray auroral emission by thick target bremsstrahlung emission. 
The same electron population could explain the radio data.

{The subject of this paper, HR\,5907, is} 
the twin of HR\,7355. 
Both stars have almost the same spectral type and {similar rotation periods}.
The two stars basically differ only by the 
{magnetic topology, that presumably affect their magnetospheres.}
Similar to HR\,7355, the X-ray spectrum of HR\,5907 gives an indication of the existence
of non-thermal X-ray radiation, which provides constraints for
the model free parameters that control the stellar radio emission.  
The opportunity to measure and model the radio light curves of HR\,5907
at many frequencies give us the capability to probe 
the physical conditions of the magnetosphere
and to obtain hints of its magnetic field topology.

In Section~\ref{info} we summarize the properties of HR\,5907 that
are already known.  Section~\ref{sec:obs} describes the radio
(VLA), millimeter (ALMA), and X-ray ({\em Chandra}) observations.  
The radio/millimeter and X-ray properties of HR\,5907 are developed 
in Sections~\ref{sec_radio} and \ref{sec_xray}.  
The results from modeling of the
HR\,5907 radio emission is detailed in Section~\ref{mod_radio}.
The effect on the radio emission of an equatorial {thin ring} of cold
thermal plasma, within the magnetosphere of HR\,5907, is
presented in Section~\ref{cold_torus}.  Section~\ref{radio_aurora}
outlines an alternative scenario that
involves the possible contribution of the auroral radio emission from HR\,5907.
The effect of a magnetic field topology more complex than a simple dipole for
the radio/millimeter emission is discussed in Section~\ref{hr5907_mag}.  
The results of this work are summarized in Section~\ref{sum}.

\begin{table}
\caption[ ]{HR\,5907  stellar parameters.}
\label{par_star}
\footnotesize
\begin{tabular}[]{lccr}
\hline
Parameter                                                      & Symbol                         &                                           &ref. \\                 
\hline
Spectral and Peculiarity Type                    &$Sp$                                 & B2.5V                                   & {1} \\
Distance [pc]                                                 &$D$                                    & $131\pm6$                                      & {2}  \\
Reddening [mag]                                         &$E(B-V)$                           & 0.14                                       & {3}  \\
{Mass  [M$_{\odot}$]}                             &{$M_{\ast}$}                & {$5.5\pm 0.5$}                                         & {3}  \\
Equatorial Radius [R$_{\odot}$]               &$R_{\ast}$                         & $3.1\pm 0.2$                                         & {3}  \\
Effective Temperature [K]                          &$T_\mathrm{eff}$          & $17000\pm1000$                                  & { 3}   \\
Rotational Period [days]                            &$P_\mathrm{rot}$              & $0.508276^{+0.000015}_{-0.000012} $                           & {3}\\
Polar Field [Gauss]                                     &$B_\mathrm{p}$                 & $15700^{+900}_{-800}$                                 &  {3} \\
Rot. Axis Inclination [degree]           &$i$                                        & $70\pm10$                                         & {3} \\
Mag. Axis Obliquity [degree]            &$\beta$                                 &  $7^{+1}_{-2}$                                        & {3}     \\
\hline
\end{tabular}
\begin{list}{}{}
\item[References:]
{(1)} \citealp{hoffleit_jaschek91};
{(2)} \citealp{van_leeuwen07};
{(3)} \citealp{grunhut_etal12a};
\end{list}
\end{table}

\section{The magnetic early B-type star HR\,5907}
\label{info}

HR\,5907 (HD\,142184) is a main sequence B2.5V star, showing
evidence of photometric and spectroscopic variability with a period
of 0.508276$^{\rm d}$ \citep{grunhut_etal12a}.  HR\,5907 is also
characterized by a strong magnetic field that is variable with the same period.  
{Among the currently known} magnetic early-type stars, 
HR\,5907 has the shortest rotation period.  
The high rotation speed makes 
{this star oblate, in fact the ratio between polar and equatorial radii is $\approx 0.88$ \citep{grunhut_etal12a}}.
HR\,5907's variability has been explained in the framework of the
Oblique Rotator Model (ORM), where the star is characterized by a
mainly dipolar magnetic field topology, having the magnetic dipole
axis tilted with respect to the rotational axis \citep{babcock49,stibbs50}.

The longitudinal magnetic field component of HR\,5907, given by the
average over the whole visible disc of the magnetic field vector
components along the line of sight, 
also defined as the effective magnetic field ($B_{\mathrm {e}}$), 
ranges from $-2000$ to $-500$ Gauss \citep{grunhut_etal12a}.  
The $B_{\mathrm {e}}$ values were
derived from the analysis of the Stokes~$V$ stellar spectrum by using
the Least-Squares Deconvolution (LSD) technique \citep{donati_etal97}.
The main stellar parameters of HR\,5907 are listed in Table~\ref{par_star}.

The magnetic curve of HR\,5907 was modeled assuming a simple dipole
(polar magnetic field strength $B_{\mathrm p} =15.7$ kG) slightly
misaligned with respect to the stellar rotation axis ($\beta=7^{\circ}$).
On the other hand, 
{the Bayesian analysis of the Stokes~$V$ profile
gives the indication of a lower polar field strength ($10.4^{+0.28}_{-0.35}$ kG) \citep{grunhut_etal12a}.
The authors also noted considerable differences between the $B_{\mathrm p}$ values
obtained by the line profile fitting process of the individual observations
(range 4--25 kG, dipole axis still almost aligned with the rotation axis).
This suggests a surface magnetic field with contributions from higher-order multipolar components.}

HR\,5907 has detectable radio emission in the centimeter/millimeter frequency range.   
The star is listed in the NRAO VLA Sky Survey (NVSS) \citep{condon_etal98}, 
with a flux density at 1.4 GHz ($\lambda \approx 20$ cm) of $12.6 \pm 0.6\,{\rm mJy}$, 
and was also detected in the AT20G survey \citep{murphy_etal10} 
using the Australian Telescope Compact Array.  
The fluxes of HR\,5907 reported by the AT20G survey are respectively: 
$46 \pm 3$ mJy at 5 GHz ($\lambda \approx 6$ cm), $95 \pm 5$ mJy at 8 GHz ($\lambda \approx 4$ cm), 
and $104 \pm 6$ mJy at 20 GHz ($\lambda \approx 1.5$ cm).  
HR\,5907 was further detected by the Korean VLBI Network (KVN) at 43 GHz ($\lambda \approx 7$ mm) \citep{petrov_etal12}.
The reported 43~GHz fluxes measured by the KVN are $140\pm21$ mJy
and $123\pm20$ mJy, as a function of the projected baseline length.
The two measured fluxes are consistent within the reported errors,
indicating that HR\,5907 is unresolved at the resolution of the KVN 
($\approx 5$ mas).

\section{Observations and data reduction}
\label{sec:obs}

Radio observations of the magnetic early-type stars are currently
limited mainly to centimeter wavelengths.  Magnetic star spectra
at the millimeter range are largely unknown.  
To our knowledge, mm flux measurements of early-type magnetic stars have been reported
only by \citet{leone_etal04}.  Those results were obtained with the
IRAM interferometer at 87.7 GHz.  
Owing to poor signal-to-noise, only two targets were clearly detected.
In this paper, we present the combined radio 
(range of wavelengths from the centimeter to the millimeter)
and X-ray analysis of what is currently the fastest rotating hot magnetic star, HR\,5907.

\subsection{VLA}
\label{sec_vla}

The radio measurements of HR\,5907 analyzed in this paper were
performed using the Karl G.\ Jansky Very Large Array (VLA) radio
interferometer, operated by the National Radio Astronomy
Observatory\footnote{The National Radio Astronomy Observatory is a
facility of the National Science Foundation operated under cooperative
agreement by Associated Universities, Inc.} (NRAO).  
The observations were carried out during two epochs, using the full array
configuration, and cyclically varying the observing frequency.
Details regarding these radio measurements are listed in Table~\ref{VLA_log}.

The standard calibration pipeline, working on the Common Astronomy
Software Applications (CASA), was used by the NRAO staff to calibrate
the interferometer measurements. The calibrated data were imaged
using the CASA task CLEAN, providing the radio maps of each observing
scan for the Stokes~$I$ and $V$ parameters of the sky region around HR\,5907.  
The flux density of the source, for each
observing frequency and for each scan, was measured by fitting a
two-dimensional gaussian at the source position in the Stokes~$I$ and $V$ cleaned maps.  
The target is located in the center of the field.
The error of the bidimensional gaussian fitting
procedure, and the map rms, measured in a field area without other
field radio sources, were summed in quadrature to estimate the
uncertainty of the single measurement.  The radio measurements are
listed in Table~\ref{radio_data_hr5907}.  On average the
uncertainty of the total flux density measurements for each observing
band are respectively: $\approx 2 \%$ at 6 and 10 GHz; 
$\approx 3\%$ at 15 GHz; $\approx 6 \%$ at 22 GHz; 
and $\approx 8 \%$ at 33 and 44 GHz.

The array was in configuration `B' during the two observing runs of HR\,5907.
The corresponding minimum synthesized beam size was
$0.33 \times 0.15$ [arcsec$^2$], obtained at the maximum observed frequency of 44 GHz. 
The size of a Gaussian fit to HR\,5907 in the image is consistent with the synthesized beam, 
indicating that the source is unresolved at the maximum resolution obtained.

\begin{table} 
\caption{VLA observing log. Array configuration: B, Code: 15A-041.} 
\label{VLA_log} 
\begin{center} 
\begin{tabular}{ccccc} 
\hline 
$\nu$        &Bandwidth          &Epoch       &Flux cal   &Phase cal       \\ 
 (GHz)                 & (GHz)                               &                   &                 &                        \\
\hline 
6/10/15     &2                           &2015-Mar-09     &3C286        & 1522$-$2730  \\ 
22/33/44   &8                          & 2015-Mar-09     & 3C286        & 1625$-$2527  \\ 
6/10/15      &2                          & 2015-Mar-16    & 3C286        & 1522$-$2730  \\ 
22/33/44    &8                         & 2015-Mar-16     & 3C286        & 1625$-$2527  \\ 
\hline             
\end{tabular}     
\end{center} 

\end{table} 

\subsection{ALMA}
\label{sec_alma}

HR\,5907 was included in the observatory's ``Weak Source Survey" (WSS) 
for potential gain calibrators \citep[][\S4.3]{ALMA_LBC_overview}, 
and was observed on 2015-Aug-02. 
HR\,5907 was included in the survey based on the previously mentioned 20~GHz 
flux density of $104\pm6\,{\rm mJy}$ \citep[AT20G,][]{murphy_etal10}. 
The WSS was designed as a dual band survey, 
observing each source with consecutive scans covering 87--103 and 277--293~GHz 
(Bands 3 and 7, respectively, 
using a common photonic LO frequency of 95~GHz for both bands). 
The photometry from the survey is available publicly in the ALMA Source Catalogue, 
which is served from the ALMA Science Portals. Due to its strong variability, 
HR\,5907 has then been flagged in the catalogue as an unsuitable calibrator. 
The measurements for HR\,5907 can be retrieved from 
{the link reported in footnote}\footnote{\url{https://almascience.eso.org/sc/rest/measurements?catalogue=5&name=J1553-2358}}. 
In this case the observation was made with the ACA correlator using 11 7m and 3 12m antennas.

HR\,5907 was observed in two pairs of scans separated by 39 minutes. 
We have re-reduced the data\footnote{ASDM UID {\tt uid://A002/Xa73e10/X3061}} to produce fluxes per scan. 
{The ALMA receivers are able to detect the linear orthogonal polarizations (X and Y) of the electromagnetic wave.
The correlation products of the X and Y components 
can be used to build the circular polarization state of the wave.}
Due to a combination of two technical issues that independently affected linearity correction 
in three of four basebands and all correlation products with the Y polarization, 
only one baseband (2~GHz bandwidth) and the XX correlation product have been used. 
{Due to these issues, the circular polarization information of the electromagnetic wave measured by ALMA could not  be extracted.
The centre frequencies of the two basebands were 102.04 and 291.97~GHz in the Band 3 and 7}, respectively. 
The corresponding wavelengths are respectively $\lambda \approx 3$ and $1\,{\rm mm}$.
The ALMA measurements are listed in Table~\ref{alma_data_hr5907}. 
An absolute flux calibration uncertainty of 5\% has been added in quadrature with the noise-based uncertainty.

\begin{figure}
\resizebox{\hsize}{!}{\includegraphics{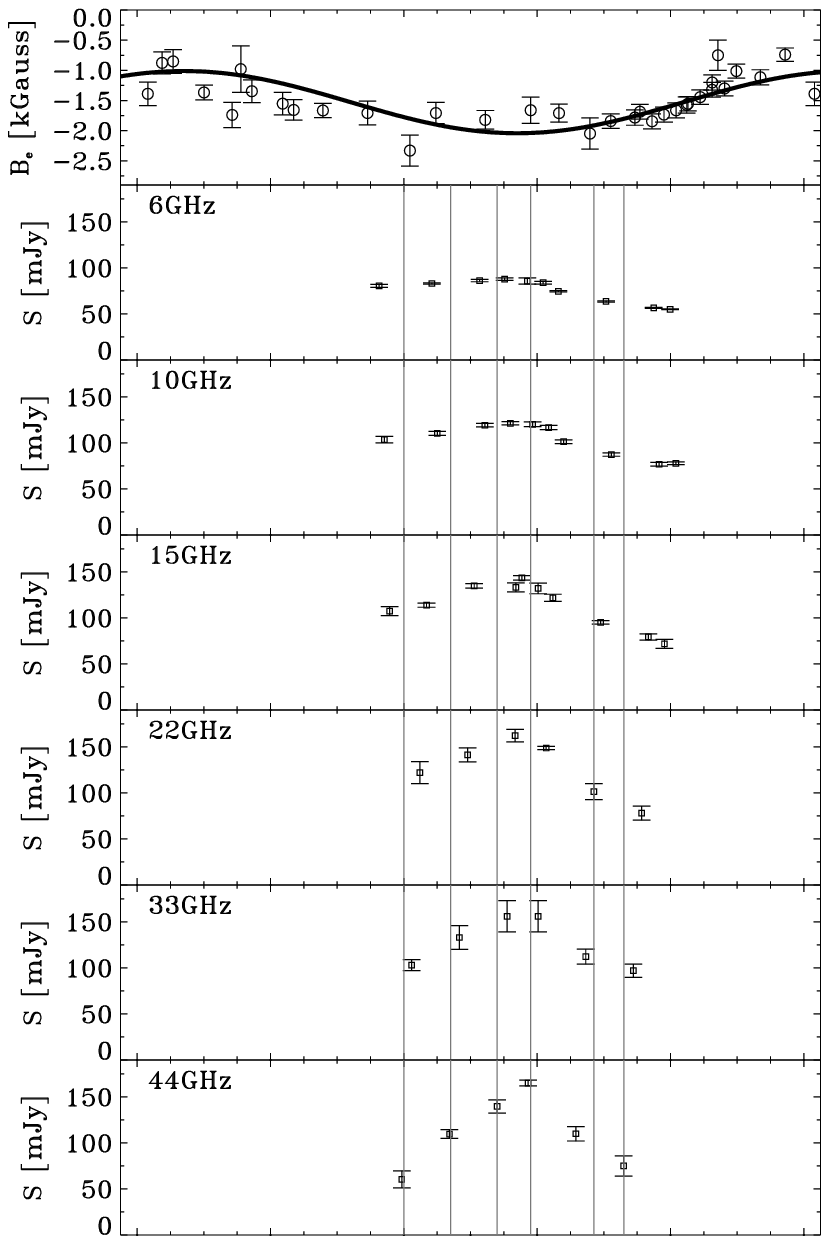}}
\resizebox{\hsize}{!}{\includegraphics{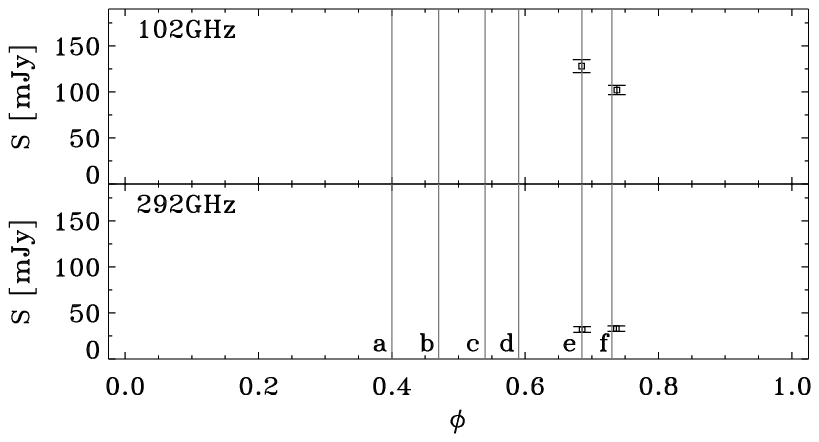}}
\caption{Multi wavelength radio light curves of HR\,5907. The measurements refer to the total intensity radio emission,
Stokes~$I$ parameter. In the top panel the stellar magnetic curve of the effective magnetic field
is also displayed.}
\label{fig_data}
\end{figure}

\begin{figure}
\resizebox{\hsize}{!}{\includegraphics{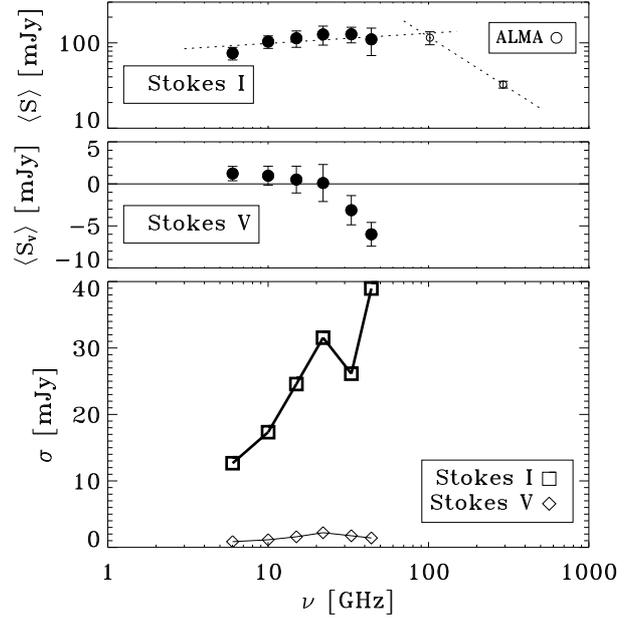}}
\caption{Average radio spectrum of HR\,5907. Top panel: total intensity radio emission, Stokes~$I$.
The two ALMA measurements are also displayed (open circles);
the filled circles refer to the VLA average measurements. 
The superimposed dotted lines indicate the
linear fit of the average HR\,5907 spectrum in the two frequency ranges, 
respectively at $\nu < 102$ GHz; and at $\nu > 102$ GHz.
Middle panel: circularly polarized radio emission, Stokes~$V$.
The error bars for the are the Stokes~$I$ and $V$ are the standard deviation of all the measurements obtained at the same radio frequency.
Bottom panel: standard deviations $\sigma$ calculated respectively for Stokes~$I$ and $V$.}
\label{sigi_sigv}
\end{figure}

\begin{figure*}
\resizebox{\hsize}{!}{\includegraphics{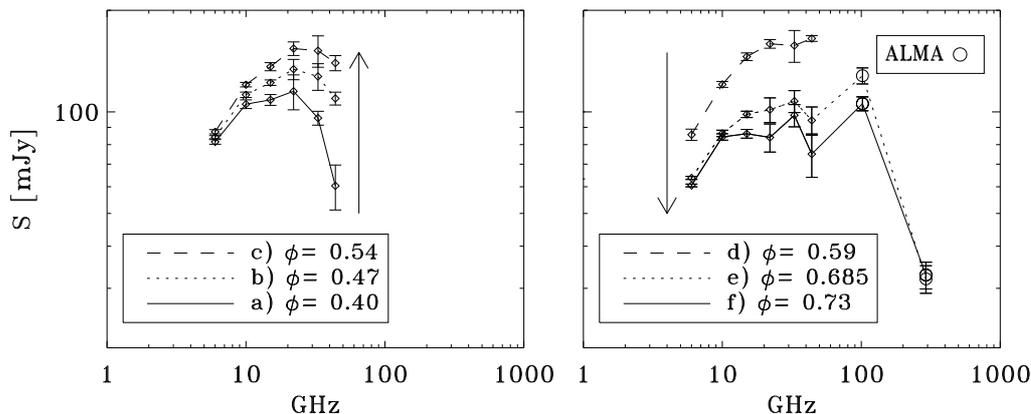}}
\caption{Spectra of the total intensity of the HR\,5907 radio emission.
The left panel displays the spectra collected 
at phases preceding the maximum flux density measured at 44 GHz.
The sense of the long arrow indicates the temporal sequence of the displayed spectra.
Right panel displays the spectra measured during the
decaying phases, and like the left panel, the sense of the arrow indicates the temporal sequence of the spectra. 
The two ALMA measurements of the total intensity at 102 and 292 GHz are also shown (open circles).}
\label{fig_spectra}
\end{figure*}

\subsection{\em Chandra}
\label{sec_x}

HR\,5907 was observed by {\em Chandra} on 2012-06-18 for 25~ks
(ObsID 13624, \citealp{naze_etal14}).  
We retrieved and analyzed these archival X-ray data. 
The ACIS-S detector was operated in the ``very faint" mode.  
Data were retrieved and reduced using the most recent calibration files.  
{The spectrum was extracted using standard
procedures from a region with diameter $\approx 3\arcsec$. 
The spectrum was grouped such that there is at least 10 counts in each energy bin.}
The background area was chosen in a nearby area free of X-ray sources.
To analyze the spectra we used the standard X-ray
spectral fitting software {\sc xspec} \citep{arnaud96}. 
The abundances were set to solar values according to \citet{asplund_etal09}.  
The adopted distance to the star and interstellar reddening $E(B - V)$
are listed in Table~\ref{par_star}.

\section{The radio/millimeter emission of HR\,5907}
\label{sec_radio}

\subsection{The multi-wavelength light curves} 
\label{sec_light_curves}

HR\,5907 was detected by the VLA and ALMA interferometers in each observing band.
The measurements were phase folded using the ephemeris 
for heliocentric Julian date as given by \citet{grunhut_etal12a}:
\begin{equation}
\label{effemeridi}
{{\mathrm {HJD}}=2~447~913.694(1)+0.508276(^{+15}_{-12})E ~{\mathrm{(days)}}}
\end{equation}

\noindent displayed in Fig.~\ref{fig_data} 
and listed in Tables~\ref{radio_data_hr5907} and \ref{alma_data_hr5907}.  
The reference heliocentric Julian date (HJD$_0$) refers
to the maximum of the H${\alpha}$ equivalent width.  
For comparison with the multi-wavelength radio light curves of the total intensity
({Stokes~$I=RCP+LCP$,} 
respectively Right plus Left Circular Polarization state),
the magnetic field curve of HR\,5907 (ORM parameters listed in Table~\ref{par_star}) 
is also shown in the top panel of Fig.~\ref{fig_data} 
superimposed to the $B_{\mathrm e}$ measurements \citep{grunhut_etal12a}.

The radio measurements were collected by the VLA in two different observing
epochs separated by about 7 days ($\approx 14$ stellar rotations).
Measurements from the two sets of data performed at the same
rotational phases merge perfectly. Even though the stellar rotational
period was not fully covered at the VLA observing bands 
(fraction of covered phases $\approx 50\%$), 
HR\,5907 exhibits a clear rotational modulation of its radio emission.  
The amplitudes of the radio emission variability grow
as the observing frequency increases.  This is made more clear when looking
at the spectral dependence of the 
{standard deviations} ($\sigma$) of
the measurements performed at the same frequency (see bottom panel of Fig.~\ref{sigi_sigv}).  
In this analysis the statistical parameter
$\sigma$ was used as index of the flux density variability.
This analysis has been performed for the VLA data only, 
because the standard deviation of the measurements 
at the two ALMA observing bands {are unreliable} as variability 
{indices} due to the { small number of measurements} available.

The temporal variations of radio/millimeter light-curves of HR\,5907 
{can be explained} by the gyro-synchrotron emission from a population of 
mildly relativistic electrons moving inside the stellar magnetosphere.
This model explains the enhancement of the measured radio emission 
when the stellar rotational phases are close to the maxima of the effective 
magnetic field curve, as commonly observed in the hot magnetic stars
\citep{trigilio_etal04,bailey_etal12,leto_etal06,leto_etal12,leto_etal17a}.

The measured radio emission at 44 GHz shows a clear
maximum related with the most negative extremum of the stellar
magnetic curve ($B_{\mathrm e} \approx -2000$ Gauss).  At the lowest
radio frequencies (6 and 10 GHz) this maximum is also present, but less prominent.
At 102~GHz there is statistically significant variation between the two scans, 
in good agreement with the VLA measurements (see Fig~\ref{fig_data}).
Even if the two ALMA measurements performed at 102 GHz show
a clear rotational variability, the poor phase sampling 
{prevents us from reaching} a reliable conclusion regarding the possible existence and
phase location of the maximum detected at the lowest frequencies.
At the still higher observing frequency, 
the ALMA measurements at 292 GHz are noise-limited and do not show significant variation.

\subsection{The spectra}

To highlight the multi-wavelength temporal evolution of the
radio emission for the total intensity (Stokes~$I$) of HR\,5907, 
the source radio spectra were obtained almost at the same rotational phases.  
Due to the most rapid temporal variation of the 
radio emission at $\nu \geq 44$ GHz (see Fig.~\ref{fig_data}), 
the spectra were collected at the phases coinciding 
with the highest radio frequency measured by the VLA, 
or with the available ALMA measurements.
These selected phases are marked by the vertical lines displayed in Fig.~\ref{fig_data}.  
The flux densities at the other frequencies were obtained from interpolating the data.  
The so derived radio spectra of the total intensity are displayed in Fig.~\ref{fig_spectra}.
Unfortunately the ALMA measurements do not cover
the same large range of phases sampled by the VLA.
The spectra of HR\,5907 that include the available ALMA measurements
are the (e) and (f) spectra.
The two rotational phases sampled by ALMA
follow the phase of maximum 44 GHz emission, (d) spectrum.

The VLA dynamic spectra indicates a clear temporal evolution of the 
HR\,5907 radio spectrum in the frequency range 6--44 GHz.
In fact, as the measured radio flux density enhances, the peak
of the spectrum also moves toward the higher frequencies.
Looking at Fig.~\ref{fig_spectra} it is clear that
the spectra frequency peak ($\nu_{\mathrm {peak}}$) continuously increases
approaching to the maximum 44 GHz emission, measured at $\phi=0.59$:
spectrum (d) displayed in the right panel of Fig.~\ref{fig_spectra}.
At this particular rotational phase the radio spectrum
does not changes its slope between 6 and 44 GHz.
However, $\nu_{\mathrm {peak}}$ could be 
at higher frequency since observations at frequencies beyond 44 GHz 
are not available at this particular stellar rotational phase.
In the framework of the gyro-synchrotron radio emission mechanism  from a
co-rotating stellar magnetosphere, the frequency of the peak in the
radio spectrum is a function of the magnetic field strength of the
magnetospheric layer that mainly contributes to the emission
at that frequency \citep{dulk_marsh82}.  
The observed increase of $\nu_{\mathrm {peak}}$
suggests that the stellar rotation reveals
magnetospheric regions with quite high local magnetic field strength.

For the spectra obtained at phases following 
the 44 GHz maximum emission (spectra (e) and (f) displayed in the right panel of Fig.~\ref{fig_spectra})
the measurements at the ALMA frequencies are available.
These spectra highlight that the HR\,5907 emission at 102 GHz 
has a flux level higher than the corresponding 44 GHz emission, 
whereas the 292 GHz measurements indicate that 
the HR\,5907 emission steeply decreases approaching at the sub-millimeter wavelengths.
These ALMA data indicate that the magnetospheric layers where the 102 GHz emission is generated
have a dominant contribution to the HR\,5907 gyro-synchrotron emission,
at least for these particular rotational phases.

Radio spectra from all the
collected data were averaged and are displayed in the top panel of Fig.~\ref{sigi_sigv}.
The spectral index of the mean radio spectrum of HR\,5907 is quite flat 
(spectral index $\approx 0.1$, estimated between 6 and 102 GHz).  
The spectral index calculated within the two ALMA measurements (102 and 292 GHz)
is instead very steep (spectral index $\approx -1$).  

Scaling for the stellar distance ($D=130$ pc), the absolute radio
luminosity of HR\,5907 is about $2 \times 10^{18}$ [erg s$^{-1}$Hz$^{-1}$]. 
This was calculated in the frequency range where the
source spectrum is almost flat averaging all the radio
flux density measurements performed between 6 and 102 GHz.  
This radio luminosity makes HR5907 about 2 times brighter than either 
$\sigma$\,Ori\,E \citep{linsky_etal92} or HR\,7355 \citep{leto_etal17a}.  
At the present HR\,5907 has the highest published luminosity 
in the cm-mm band for a non-degenerate magnetic massive star.

\subsection{The circularly polarized emission}
\label{circ_pol}

\begin{figure}
\resizebox{\hsize}{!}{\includegraphics{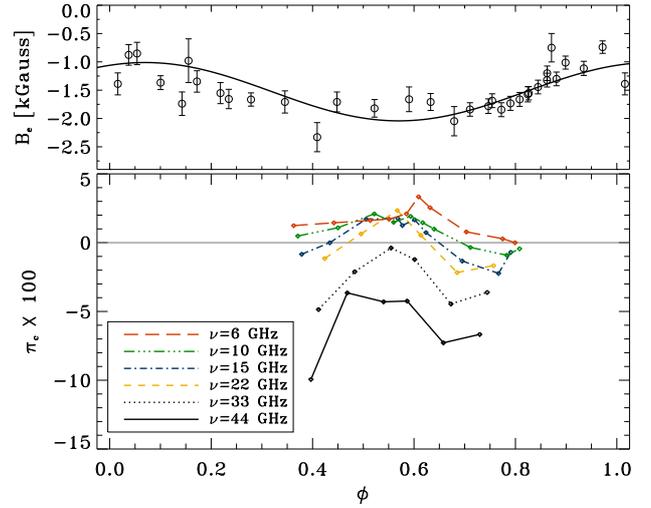}}
\caption{
Bottom panel: fraction of the circularly polarized emission 
($\pi_{\mathrm c}={\mathrm {Stokes}~V}/{\mathrm {Stokes}~I}$)
as a function of stellar rotational phase. 
The light curves of $\pi_{\mathrm c}$ were calculated for each observing frequency.
Top panel: effective magnetic field curve, as displayed in the top panel of Fig.~\ref{fig_data}, 
shown here for reference. 
}
\label{fig_datav}
\end{figure}

The fractional circular polarization 
($\pi_{\mathrm c}={\mathrm {Stokes}~V}/{\mathrm {Stokes}~I}$, 
where Stokes~$V=RCP-LCP$, respectively Right minus Left Circular Polarization)
is also variable for HR\,5907. 
The values of $\pi_{\mathrm c}$ as a function of the rotational phase are shown in the bottom panel of Fig.~\ref{fig_datav}; 
in the top panel the magnetic curve of HR\,5907 is shown for reference.  
The error bars of the $\pi_{\mathrm c}$ 
measurements were not displayed so that the rotational modulation 
of the fraction of the circular polarization could be seen more clearly.  
The average uncertainties of $\pi_{\mathrm c}$ for each observing frequency are respectively: 
0.3\% at 6 GHz, 0.2\% at 10 GHz, 0.3\% at 15 GHz, 0.4\% ant 22 GHz, 0.6\% at 33 GHz, and 1.1\% at 44 GHz.  
The estimated error levels are low enough to consider the amplitudes of 
the $\pi_{\mathrm c}$ rotational modulations observed at each radio frequency as significant.
{In fact the observed amplitudes of the $\pi_{\mathrm c}$ variations are from $\approx 6$ to 14 times higher than the uncertainties.}

HR\,5907 shows a frequency-dependent effect of the measured fraction
of the circularly polarized radio emission. The sign of the polarized
radio emission changes as the radio frequency increases.  At the
sampled rotational phases, the sign of $\pi_{\mathrm c}$ is always
positive at 6 GHz, but $\pi_{\mathrm c}$ reverses its sign 
at intermediate frequencies ($\nu=10$, 15, and 22 GHz) as the star rotates.  
On the other hand,  $\pi_{\mathrm c}$ is always negative at higher frequencies
($\nu=33$ and 44 GHz); see the bottom panel of Fig.~\ref{fig_datav}.
The VLA measurements highlight that the absolute magnitude of the
polarized emission is stronger at the highest observing frequency ($\nu= 44$ GHz).

The overall behavior of the circularly polarized emission of HR\,5907
can be analyzed averaging all the measurements performed at the same frequency. 
The variability index $\sigma$ calculated for the
Stokes~$V$ emission (also shown in the bottom panel of Fig.~\ref{sigi_sigv})
indicates that the amplitude of the Stokes~$V$ rotational modulation
does not significantly vary with radio frequency,
in contrast to what is observed for Stokes~$I$.  
The Stokes~$V$ mean spectrum is displayed in the middle panel of Fig.~\ref{sigi_sigv}.
The frequency-dependent behavior of the circularly polarized emission
of HR\,5907, already mentioned by analyzing the light curves of $\pi_{\mathrm c}$,
is now more clearly highlighted.  On average the circularly
polarized emission at 6 and 10 GHz is positive, at 15 and 22 GHz
is null, and at 33 and 44 GHz is negative.  
This indicates that, on average, the left-handed circularly
polarized radio emission of HR\,5907 prevails at high frequencies, 
whereas at the low frequencies the right-handed circularly polarized emission 
is slightly stronger.

In the case of the gyro-synchrotron emission mechanism, the
polarization sign relates to the average spatial orientation of the
magnetic field vectors in the regions where the radio emission at
a given frequency mainly originates.  
Consequently the circularly polarized radio emission 
and the effective magnetic field sign are correlated.  
When the northern magnetic hemisphere (magnetic field
lines mainly oriented toward the observer) is visible, the radio
emission shows right-handed circular polarization (Stokes~$V$ positive).
On the other hand the radio emission is left handed circularly polarized 
(Stokes~$V$ negative) when the southern hemisphere is visible. 
This correlation was clearly observed in the cases of CU\,Vir \citep{leto_etal06},
$\sigma$\,Ori\,E \citep{leto_etal12}, and HR\,7355 \citep{leto_etal17a}.
that are all three well-known hot magnetic stars characterized by
a dominant dipolar magnetic field and well described by the ORM,
{even if the magnetic curve shape
of CU\,Vir \citep{kochukhov_etal14} and $\sigma$\,Ori\,E \citep{oksala_etal12,oksala_etal15} evidence 
that these stars possess a quadrupolar component that is not-negligible at the stellar surface.}

In the case of HR\,5907 the circularly polarized emission at $\nu
\ge 33$ GHz is left handed circularly polarized, in accordance with
the measured sign of the effective magnetic field.  The polarization
sign of the radio emission at the lower frequencies is instead
anti-correlated with respect to the sign of $B_{\mathrm e}$, as
clearly shown in Fig.~\ref{fig_datav}.

\section{The X-ray emission of HR\,5907}
\label{sec_xray}

HR\,5907 is a known relatively hard X-ray source which was 
detected by the {\em ROSAT All Sky Survey} \citep{oskinova_etal11}, 
and more recently observed {by the {\em Chandra} X-ray telescope}
(see Section\,\ref{sec_x}).  
{{\em Chandra} exposure time covered about a half of stellar rotation period. 
To check for possible X-ray variability, we extracted an X-ray light curve in 0.3--10.0\,keV energy band. 
Statistical tests did not reveal X-ray variability or systematic changes in the X-ray light curve. 
New, longer observations are required to fully investigate possible X-ray variability associated with stellar rotation.
}

Similar to its twin, HR\,7355, HR\,5907 strongly violates the empiric 
Gu\"{e}del-Benz relation between the X-ray and radio luminosities of 
coronal main sequence stars  
$L_{\mathrm X} / L_{\nu, {\mathrm {rad}}} \approx 10^{15.5}$\,Hz \citep{{guedel_benz93,benz_guedel94}}. 
{The above relation holds for stars covering a wide Sp range, from the F to the early type M.}
For HR\,5907, this ratio is more than 3 orders of magnitude lower, highlighting that the physical 
processes responsible for the radio and {X-ray generation} in magnetic B-type 
stars are distinct from the mechanisms working in the coronae of the 
{late-type stars.}

\begin{figure}
\resizebox{\hsize}{!}{\includegraphics[angle=-90,origin=c]{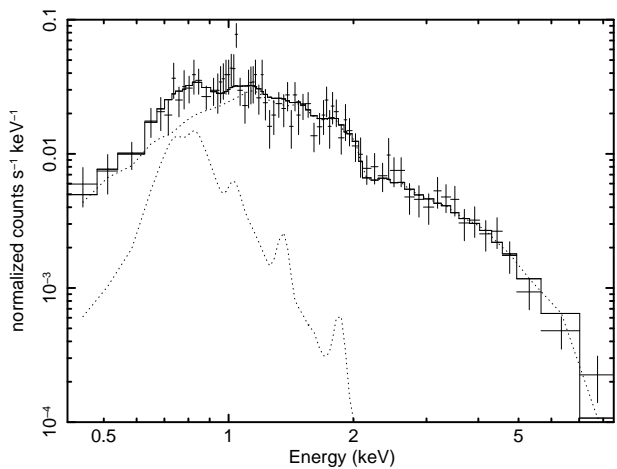}}
\caption{{\em Chandra}  ACIS-S spectrum of HR\,5907 with error bars corresponding
to 3$\sigma$ with the best fit thermal ({\it apec}) plus power-law model (solid line). 
{The dashed lines show separate contributions of thermal and power-law spectral models}.
The model parameters are shown in Table~\ref{par_xray}. 
 }
\label{fig_xray}
\end{figure}

{The {\em Chandra} ACIS-S} spectrum of HR\,5907 is shown in 
Figure\,\ref{fig_xray}. It could be well fitted {(reduced $\chi^2 = 0.67$)}  
by a two-temperature thermal plasma model \citep[see also][]{naze_etal14}. 
{The temperature of the hottest plasma component is well 
constrained ($6.5\pm 0.9$\,keV or $75 \pm 10$\,MK, Table~\ref{par_xray}).
It is clear that the presence of such hot plasma cannot be explained in the 
framework of the MCWS model. In fact,
from} the Rankine-Hugoniot condition for the strong shock ($T {\rm [MK]} 
\approx 14 ~ (v_{\mathrm W} / 10^3 {\mathrm {km}}~{\mathrm s}^{-1})^2$), 
to reach the temperature of 
{75\,MK the wind of HR\,5907 should have velocity of $\approx 2300$\,km~s$^{-1}$.
Even if the fast rotation can increase the wind velocity \citep{bard_townsend16}, 
the above estimated velocity}
is not physical for a B2.5V 
star -- empirically estimated terminal wind velocities of the main sequence 
B-type stars do not exceed 1000\,km\,s$^{-1}$ \citep{prinja89,oskinova_etal11}.
Such low wind velocity is sufficient to explain only the presence of the cooler plasma component 
{($(0.45\pm 0.18)$\,keV or $(5 \pm 2)$\,MK, Table~\ref{par_xray}).}

As a next step, by analogy with HR\,7355, we fitted the X-ray spectrum of 
HR\,5907 using a spectral model that combines thermal and non-thermal 
(power-low) components. The quality of fit is 
{similar to the two temperature model (reduced $\chi^2 = 0.67$)}. 
Figure ~\ref{fig_xray} shows the 
observed and the fitted model, while the model parameters are listed in 
Table~\ref{par_xray}. In this spectral model, the thermal plasma component 
has a temperature of {$\approx 12$ MK (1\,keV), 
follow that the wind velocity is plausible for a B2.5V star ($\approx 900$\,km\,s$^{-1}$,}
using the Equation~10 reported in \citealp{ud-doula_etal14}).

\begin{table}
\begin{center}
\caption[ ]{The X-ray spectral parameters derived from the {\em Chandra} ACIS-S spectrum of HR\,5907 
assuming a two-temperature CIE plasma ({\em apec}) model and 
thermal plus power-law model. 
The models were corrected for the interstellar absorption (using {\em tbabs} model).
The spectral fits corresponding to the thermal plus power-law model are shown in Fig.~\ref{fig_xray}.
}
\label{par_xray}
\footnotesize
\begin{tabular}[]{lc}
\hline
\hline
$N^{\mathrm a}_{\mathrm H}$ [$10^{21}$ cm$^{-2}$] & $1.4\pm 0.4$                   \\
\hline
\multicolumn{2}{l}{Two temperature thermal model}\\
\hline

$kT_1$ [keV]                                                 &$0.45\pm0.18$   \\
$EM_1$ [$10^{51}$ cm$^{-3}$]                                &$4.5\pm 4.0$   \\
$kT_2$ [keV]                                                                 &$6.5\pm 0.9$   \\
$EM_2$ [$10^{51}$ cm$^{-3}$]                                 &$58 \pm 3$   \\
$\langle k T \rangle \equiv \sum_i k T_i \cdot EM_i / \sum_i EM_i$ [keV]       &6.1   \\
Flux$^{\mathrm b}$ [$10^{-13}$ erg cm$^{-2}$ s$^{-1}$]                                                    &6.0   \\
\hline

\multicolumn{2}{l}{Thermal plus power-law ($A(E)=KE^{-\alpha}$) model}\\

\hline

$kT_1$ [keV]                                                                                              &$1.0\pm0.1$   \\
$EM_1$ [$10^{51}$ cm$^{-3}$]                                                            &$3.5 \pm 1.2$   \\
$\alpha$                                                                                                     &$1.6\pm0.1$   \\
$K$ [photons keV$^{-1}$ cm$^{-2}$ s$^{-1}$ at 1 keV]                    &$(7.8 \pm 0.6)\times10^{-5}$   \\
Flux$^{\mathrm b}$ [$10^{-13}$ erg cm$^{-2}$ s$^{-1}$]                &6.3   \\
\hline
$L_{\mathrm X}^{\mathrm b}$ [erg s$^{-1}$]                                      &$1.3 \times 10^{30}$   \\
$L_{\mathrm X}/L_{\mathrm {\nu,rad}}$ [Hz]                                                               & $6.5 \times 10^{11}$   \\
\hline
\end{tabular}
\begin{list}{}{}
\item[$^{\mathrm{a}}$] exceeds the ISM hydrogen column density
\item[$^{\mathrm{b}}$] dereddened; in the 0.2--10 keV band
\end{list}
\end{center}
\end{table}

{Formally, the X-ray spectrum of HR\,5907 can also be well fitted by the 
power-low model only ($\alpha=1.8\pm 0.1$, reduced $\chi^2 = 0.8$) or single 
temperature model only ($kT=6.2\pm 0.8$\,keV, reduced $\chi^2 = 0.81$).} 
Nevertheless, we favor the combined, 
thermal and non-thermal plasma model as better motivated physically.  

By analogy with HR\,7355, the non-thermal X-rays in the HR\,5907 spectrum could 
be explained as bremsstrahlung emission from a non-thermal electron population. 
These non-thermal electrons are also responsible for the 
gyro-synchrotron stellar radio emission. When they impact the stellar 
surface,  X-rays are radiated by thick-target bremsstrahlung emission.
This physical process is well understood, in particular, the spectral 
index $\alpha$, of the non-thermal photons, can be related to the spectral 
index $\delta$, of the non-thermal electron population, by the simple relation 
$\delta=\alpha +1$ \citep{brown_71}. Then, using 
{the best fit to HR\,5907's X-ray spectrum,}
the spectral index of the non-thermal electron energy 
distribution is  $\delta=2.6$.

\begin{table*}
\caption[ ]{List of the free parameters of HR\,5907}
\label{mod_par}
\begin{tabular}{lcclc}
\hline
   &Symbol         &Range          &Simulation step &Number of simulations\\
\hline
Alfv\'en radius [R$_{\ast}$]                                &$R_\mathrm{A}$      &$8$ -- $20$            &$\Delta R_\mathrm{A}=1$ & 13\\
Thickness of the middle magnetosphere [R$_{\ast}$]                     &$l$                 &$0.1$ -- $20$          &$\Delta \log l\approx0.1$ &23 \\
Non thermal electron density [cm$^{-3}$]                              &$n_\mathrm{r}$      &$10^2$ -- $10^{5}$          &$\Delta \log n_\mathrm{r}\approx 0.1$  &30\\
Thermal electron density at the stellar surface [cm$^{-3}$]  &$n_\mathrm{0}$    &$10^{8}$ -- $10^{13}$  &$\Delta \log n_\mathrm{p_0}\approx0.25$ &20\\
\hline
\end{tabular}
\end{table*}

Interestingly, a similar mechanism is also acting in the case of the 
magnetized planet -- Jupiter. The continuously injected non-thermal electrons, 
with energies higher than $\approx 100$ keV, precipitate 
{towards the Jovian atmosphere} giving rise to auroral X-ray emission recognized in the 
{\em XMM-Newton} X-ray spectrum of Jupiter 
\citep{branduardi-raymont_etal_07,branduardi-raymont_etal_08}.

We suggest that the non-thermal X-ray emission of HR\,5907 is 
{also of auroral origin, whereas the thermal plasma component with 
the temperature $\approx 12$~MK is produced via MCWS mechanism.
This is the scenario already proposed for HR\,7355. 
For clarity, this is summarized again in the Appendix~\ref{appendix_mag_hot_star} and pictured in Fig.~\ref{modello}.}

\section{Modeling HR 5907's Radio Emission}
\label{mod_radio}

\subsection{The model}
\label{scenario}

In order to investigate the physical conditions of the magnetosphere
of a typical hot magnetic star, we developed a model of
its non-thermal radio continuum emission \citep{trigilio_etal04,leto_etal06}. 
The model was developed under the simplified hypothesis 
of a magnetic topology described by a dipole field, 
in the framework of the ORM.

Accordingly to the MCWS model, 
the stellar wind is channeled by the closed field lines and accumulates at low magnetic latitudes,
``inner-magnetoshere".
As a first approximation, the temperature of the plasma linearly increases as
a function of radial distance from the star, whereas density decreases
outward as $r^{-1}$ \citep{babel_montmerle97}.  
In our model, the plasma temperature at the 
stellar surface was assumed equal to the effective temperature ($T_{\mathrm {eff}}$).
The resulting plasma pressure can be assumed as roughly constant within the inner-magnetosphere.

The spatial surface that locates the region where the magnetic energy density equals
the plasma one is the Alfv\'{e}n surface.
The last closed magnetic field line intercepts the magnetic equatorial plane at the Alfv\'{e}n radius ($R_{\mathrm A}$).
In the case of a simple dipole, the magnetic field line is defined by the equation: $r = R_{\mathrm A} \cos^2 \lambda$
(where $\lambda$ is the magnetic latitude).
The size of the magnetospheric cavity (``middle-magnetosphere'') 
located just outside the Alfv\'{e}n surface, 
where the non-thermal electrons are efficiently generated, 
is defined by the length ($l$) of the acceleration region.
In our magnetospheric model,
the density of the thermal plasma within the middle-magnetosphere
was assumed equal to the density of the freely escaping stellar wind.

The relativistic electrons are characterized by a power-law energy
spectrum ($N(E) \propto E^{-\delta}$), and the directions of their
velocity vectors are isotropically distributed with respect to the local
magnetic field vector.  
The fraction of the non-thermal electrons moving toward the stellar
surface travels through layers with increasing magnetic field strength.  
As a consequence of the balance between the reducing magnetic flux
tube section and the magnetic mirroring effect, the density of the
non-thermal electrons can be assumed constant within the
middle-magnetosphere.

The  relativistic electron population freely propagates within the
middle magnetosphere and radiates an incoherent non-thermal radio continuum
by the gyro-synchrotron emission mechanism, that
is a function of the local magnetic field strength and orientation,
and of the density of the ambient ionized medium.  For the calculation
of the gyro-synchrotron emission and absorption coefficients 
the general expressions given by \citet{ramaty69} at the
low harmonic numbers ($\nu/\nu_{\mathrm B} \leq 3$, where $\nu_{\mathrm
B} \propto B$ is the local gyro-frequency) have been used, 
whereas at the higher harmonic numbers 
the approximate expressions given by \citet{klein87} have been adopted.  
Both calculation approaches take into account
the presence of the ionized ambient medium.

The ionized medium affects the gyro-synchrotron radiation through the frequency-dependent
suppression mechanism known as the Razin effect. 
The Razin effect operates on each propagation
mode, but with a different magnitude between the ordinary (O) and
the extraordinary (X) modes. 
The radio wave propagation within the magneto-active circumstellar plasma requires,
also, knowledge of the refractive indices and of the polarization
coefficients, respectively for the ordinary (O-mode) and extraordinary
(X-mode) magneto-ionic modes \citep{klein_trotter84}.
The frequency-dependent absorption effects \citep{dulk85}  
of the thermal ionized plasma trapped within the inner-magnetosphere 
has also been taken into account.

The stellar magnetospheric volume is then sampled by using a 3D
cartesian grid, and all the physical quantities, needed for the
calculation of the gyro-synchrotron absorption and emission
coefficients, are calculated for each grid point, that is assumed
as a homogeneous radio source.  Following,
the radiative transfer equation can be numerically solved along the
directions coinciding with the line of sight; this is done separately
for the two magneto-ionic modes.  Finally, for a given magnetospheric
geometry, by varying the stellar rotational phase, we are able to
simulate the rotational modulation of the gyro-synchrotron stellar
radio emission, for the Stokes~$I$ and $V$, and the corresponding
brightness spatial distribution.  
For details see \citet{trigilio_etal04} and \citet{leto_etal06}.

\subsection{The numerical simulations}
\label{num_sim}

To reproduce the multi-wavelengths radio light curves of
HR\,5907 measured between 6 and 44 GHz, we performed an extensive
set of model simulations.  
Many of the fixed stellar parameters
required as input by the model are listed in Table~\ref{par_star}.
These are the distance of the star, the radius, the effective
temperature, the strength of the polar magnetic field, and the
geometry of the ORM. The other fixed parameters are the spectral
index ($\delta$) of the power-law energy spectrum of the non-thermal
electrons, and their low-energy cutoff.  The spectral index was
fixed using the constraint achieved by the fit of the HR\,5907 X-ray spectrum.  
From the analysis discussed in Sec.~\ref{sec_xray}, $\delta=2.6$ was adopted.  
The low-energy cutoff of the non-thermal electrons
was fixed  at 100~keV (corresponding to a Lorentz factor
$\gamma=1.2$) like the case of the near-twin star HR\,7355, that was the
subject of a similar study \citep{leto_etal17a}.

The model free parameters are the Alfv\'{e}n radius ($R_{\mathrm A}$), 
the length of the acceleration region for the non-thermal electrons,
the number density of these electrons, and the density at the stellar
surface of the thermal plasma trapped within the inner magnetosphere.
The ranges of the simulated parameters, with the corresponding
steps, are listed in Table~\ref{mod_par}.  
The value of $R_{\mathrm A}$ is a function of the radiatively driven stellar wind parameters,
in particular of the wind mass-loss rate $\dot{M}$, 
which is unknown in the case of HR\,5907.  
Once $R_{\mathrm A}$ is defined, the wind
parameters and consequently the plasma density within the
middle-magnetosphere are determined 
(see \citealp{leto_etal06} for details).

To simulate the radio emission of HR\,5907 as a function of the rotational phase, 
the simulations were performed for six different values of $\phi$.
Taking into account the variation of the rotational phase and the
values of the free parameters listed in Table~\ref{mod_par},
more than $10^6$ simulations were carried out.
For the simulations of the HR\,5907 radio emission
a three-dimensional cartesian grid with variable sampling step was adopted. 
The grid points at distances higher than 
12 R$_{\ast}$ are spaced every 1 R$_{\ast}$;
at the intermediate distance (range 8--12 R$_{\ast}$)
the grid spacing is equal to 0.3 R$_{\ast}$;
whereas close to the star (distance lower than 8 R$_{\ast}$),
a denser sampling grid was used (step 0.1 R$_{\ast}$).

\subsection{Comparison with the observations}
\label{sim_obs}

\begin{figure}
\resizebox{\hsize}{!}{\includegraphics{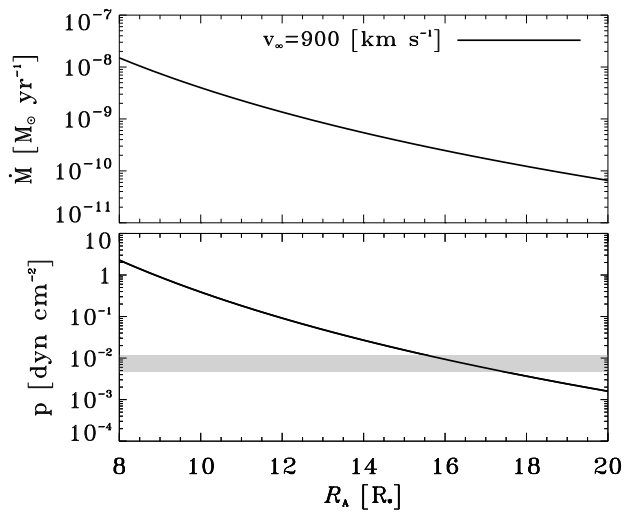}}
\caption{Top panel: The dependence of the wind mass-loss rate as a function of the Alfv\'{e}n radius.
Bottom panel: Wind ram pressure as a function of $R_{\mathrm A}$.
The shaded area locates the thermal pressure of the plasma trapped within the inner-magnetosphere
corresponding to  $n_0=$2--5 $\times 10^9$ [cm$^{-3}$].}
\label{beta}
\end{figure}

After exploring a broad space of free parameters, listed in
Table~\ref{mod_par}, we were not able to find a combination 
to reproduce simultaneously the multi-wavelength
radio light curves of HR\,5907 for both Stokes~$I$ and $V$.  
In a previous paper, using the same 3D model and the same simulation
approach, we successfully simulated the multi-wavelength radio light
curves of HR\,7355 (a twin of HR\,5907), simultaneously for the
total intensity and the circularly polarized radio emission \citep{leto_etal17a}.  
Both stars have similar radio and X-ray luminosities. 
The radio spectra of the two stars are almost flat in the spectral range observed by the VLA, 
even if the spectral slope of HR\,5907 is slightly positive ($\approx 0.1$), while the
slope measured in the case of HR\,7355 is slightly negative ($\approx -0.1$).
The average radio luminosity measured at the lowest observed frequency
(6 GHz) is quite similar for the two stars, respectively:
$L_{\nu, {\mathrm {rad}}}=1.5\times10^{18}$ [erg s$^{-1}$ Hz$^{-1}$] for HR\,5907;
$L_{\nu, {\mathrm {rad}}}=1.2\times10^{18}$ [erg s$^{-1}$ Hz$^{-1}$] for HR\,7355.  
These two stars are expected to have nearly twin magnetospheres.  
The main difference between the two stars is the dipole axis misalignment, 
that in the case of HR\,5907 is almost parallel to the rotation axis (7 degrees), 
whereas HR\,7355 is characterized by a large dipole axis misalignment
($\beta=75^{\circ}$, \citealp{rivinius_etal13}).

To take advantage of the close similarity between HR\,5907 and
HR\,7355, we exploit the free parameters that were fixed by the
HR\,7355 radio emission modeling to simulate synthetic radio
light curves corresponding for the geometry of HR\,5907.  In particular,
we derive for HR\,7355 the thermal electron density at the stellar
surface that is able to reproduce the observed rotational modulation
of its radio emission: $n_\mathrm{0}=$2--5$ \times 10^9$ [cm$^{-3}$]
\citep{leto_etal17a}.

\begin{table}
\begin{center}
\caption[ ]{Derived parameters of HR\,5907}
\label{deriv_par}
\begin{tabular}[]{lcl }
\hline
$v_{\infty}$                                              [km s$^{-1}$]                                                         &{900}                           & from the X-ray spectrum fit  \\                                                                                                                                                                     
$\delta $                                                                                                                                 &$2.6$                           & from the X-ray spectrum fit  \\                                                                                                                                                                     
$n_\mathrm{r} \times l$                    [cm$^{-2}$]                                                                &$2.8 \times 10^{16}$       & from the Stokes~$I$ at 6 GHz   \\                                                                                                                                                                     
$R_\mathrm{A}$                            [R$_{\ast}$]                                                                  &{15.5--17.5}                           & with $n_\mathrm{0}=$2--5 $\times 10^9$ [cm$^{-3}$] $^{\dag}$       \\                                                                                                                                                                                                                                                                                                                                           
$<\dot{M}>$                                       [M$_{\odot}$ yr$^{-1}$]                                               &$2.2\times10^{-10}$  &   \\                                                                                                                                                                     
$<T>$                                               [MK]                                                                                    &$0.19$                          & \\ 
$<EM>$                                               [$10^{55}$ cm$^{-3}$]                                                    &$1.8$                            & \\ 
$\dot{M}_{\mathrm {act}}$               [M$_{\odot}$ yr$^{-1}$]                                               &$6.8\times10^{-12}$   & \\               
\hline
\end{tabular}
\begin{list}{}{}
\item[$\dag$]
The thermal electron density used to simulate the HR\,5907 radio emission has been fixed 
on the basis of the results achieved with the analysis of HR\,7355 \citep{leto_etal17a}
\end{list}

\end{center}
\end{table}

\begin{figure*}
\resizebox{\hsize}{!}{\includegraphics{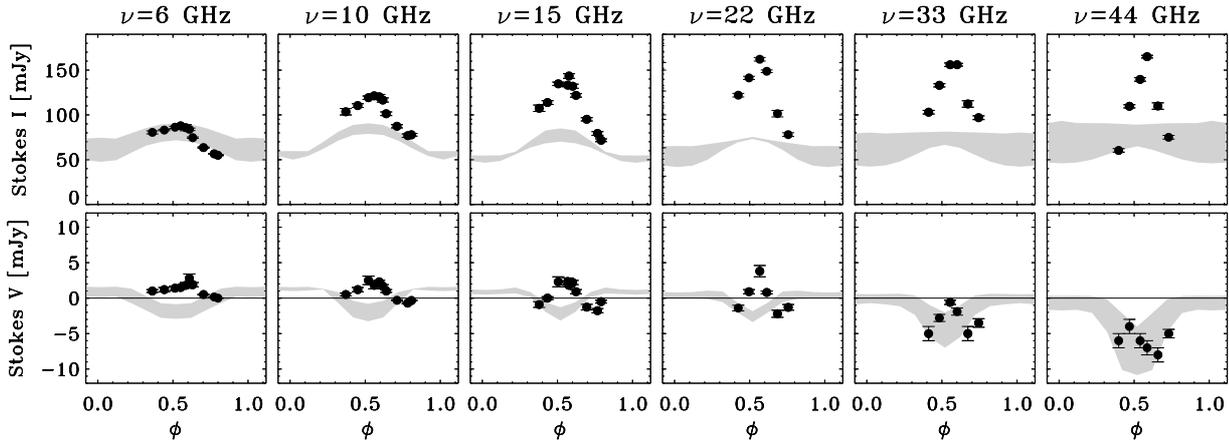}}
\caption{Top panels: multi-wavelength radio light curves of the total intensity radio emission (filled dots), Stokes~$I$,
compared with the synthetic light curves obtained using the model parameters listed in Table~\ref{deriv_par} (gray areas).
Bottom panels: the gray areas are the envelopes of the simulated light curves 
of the circularly polarized emission (Stokes~$V$) compared against
the observed ones as filled dots.
}
\label{simul_dati_ori}
\end{figure*}

In the case of a simple dipole, the equatorial value of the Alfv\'{e}n surface,
$R_{\mathrm A}$, can be located from equating the wind energy density,
radial plus centrifugal, with the magnetic energy density.  
In this simplified case, the energy balance equation can be solved
\citep{trigilio_etal04}, making it possible to directly relate the
value of $R_{\mathrm A}$ with the stellar wind parameters: 
mass-loss rate ($\dot{M}$) and terminal velocity ($v_{\infty}$).  
On the other hand, the wind parameters and the trapped plasma parameters
must be physically compatible together.  
In fact, the pressure of the thermal plasma trapped within the inner-magnetosphere must 
be lower than the wind ram pressure, so that the plasma is contained
within the Alfv\'{e}n surface deduced from the stellar wind.

The plausible terminal wind velocity is $v_{\infty}\approx 900$ km s$^{-1}$.  
Given the density of the trapped plasma at the stellar surface, the range of the Alfv\'{e}n
radii for which the wind ram pressure is compatible with the pressure
of the thermal plasma trapped within the inner-magnetosphere can
be deduced.  The dependence of the wind mass-loss rate of HR\,5907
as a function of $R_{\mathrm A}$ is displayed in the top panel of
Fig.~\ref{beta}, whereas the corresponding wind ram pressure is
pictured in the bottom panel of the same figure.  By comparing the
wind ram pressure with the pressure of the trapped thermal plasma,
the range of values of $R_{\mathrm A}$ that are physically plausible
is uniquely defined.  The so derived range of $R_{\mathrm A}$ and
the corresponding average wind mass-loss rate, temperature of the
trapped plasma, and emission measure, are listed in Table~\ref{deriv_par}.
The radiative wind from HR\,5907 can freely propagate from the
northern and southern polar caps only. 
The polar cap areas can be calculated from intersecting the magnetic field line passing through
$R_{\mathrm A}$ at the magnetic equatorial plane with the stellar surface.  
The rate of mass actually lost ($\dot{M}_{\mathrm {act}}$) from HR\,5907 
is also listed in Table~\ref{deriv_par}.

The column density of the non-thermal electrons, required to reproduce
the observed flux level at 6 GHz of HR\,5907, is 
higher than the corresponding value obtained from the simulations
of the multi-wavelength radio light curves of HR\,7355 \citep{leto_etal17a}.  
In detail, for HR\,5907 the average column density of the relativistic electrons is 
$n_{\mathrm r} \times l \approx 2.8 \times 10^{16}$ cm$^{-2}$ (listed in Table~\ref{deriv_par}),
the corresponding value for HR\,7355 is $\approx 1.8\times10^{16}$cm$^{-2}$.

The case of a reduced polar field strength \citep{grunhut_etal12a}
has been also taken into account.
We performed simulations of the gyro-synchrotron emission
from a dipole shaped magnetosphere
using the stellar parameters listed in Table~\ref{par_star},
but using $B_{\mathrm p}=10400$ Gauss \citep{grunhut_etal12a}.
In this case, to reproduce the flux level of HR\,5907 measured at 6 GHz, 
we need to enhance the column density of the relativistic electrons.
On the average the relativistic electron density has to be increased by a factor $\approx 1.5$. 
Apart this, there are no significant differences between these new simulations and 
those performed using $B_{\mathrm p}=15700$ Gauss.

The Stokes~$I$ and $V$ synthetic radio light curves of HR\,5907,
calculated using the model parameters listed in Table~\ref{deriv_par}
and the stellar parameters listed in Table~\ref{par_star},
are compared to the observed ones in Fig.~\ref{simul_dati_ori}.  
In the top panels of Fig.~\ref{simul_dati_ori}, the shaded areas
superimposed with the radio measurements represent the envelope of
the simulated radio light curves for the total intensity radio emission.  
It is clear from the figure that the simulations well
reproduce the observations only at the lower observed radio frequency (6 GHz).  
As the radio frequency increases, the discrepancy between
observations and simulations grows dramatically.  
In particular, the HR\,5907 flux level measured at $\nu \geq 10$ GHz 
is about 50\% higher than the simulated one. 
Furthermore, using the HR\,5907 geometry, 
the radio emission modulation predicted for a simple
dipole is not compatible with the observed one.  

The comparison between the simulated radio light curves for the
circularly polarized emission with the observed ones 
(bottom panels of Fig.~\ref{simul_dati_ori}) 
demonstrates that the field topology for HR\,5907,
derived from the effective magnetic field measurements, 
is able to well reproduce the left-handed circularly polarized emission measured at 44 GHz, 
roughly reproduce the average flux level of the Stokes~$V$ emission at 33 GHz, 
and totally fails to reproduce the Stokes~$V$ emission at the lower frequencies. 
In particular, regarding the comparison of the simulations with 
the measured Stokes~$V$ emission at $\nu < 33$ GHz, 
we observe that the model simulates circularly
polarized radio emissions of opposites sign with respect to the observed ones.

In summary, the model of the HR\,5907 magnetosphere is not able to
reproduce the high-frequency flux level  measured at certain stellar
rotational phases, the wide rotational modulation of the radio light
curves of the total intensity at $\nu \geq 10$ GHz, nor the sign
of the circularly polarized emission at the lowest frequencies.
In the next sections we explore possible additional effects that
could help to explain the discrepancy between the model and the
observations.

\section{Discussion}
\subsection{Effect of the cold plasma around HR\,5907}
\label{cold_torus}

{Similarly to HR\,7355, the magnetosphere of HR\,5907 is likely to trap clouds of dense and cold material}
\citep{grunhut_etal12a,rivinius_etal13}.
For both stars, these plasma clouds have 
linear extensions comparable with their stellar radii.  
Hence, the stellar magnetospheres might be 
more inhomogeneous than has been assumed in the model (Sec.~\ref{scenario}).  
{In particular, the RRM semi-analytic model of \citet{townsend_owocky05} 
predicts the formation of an equatorial thin warped disk
rigidly co-rotating with the star, extending beyond the Kepler coronation radius for several stellar radii. 
The Kepler radius (defined by the relation: $R_{\mathrm K}=(G M_{\ast} / \omega^2)^{1/3}$,
where $G$ is the gravitational constant, $\omega$ the angular velocity of the star, and $M_{\ast}$ the stellar mass,
whose value for HR\,5907 is reported in Table~\ref{par_star})
locates the equatorial region where the outward centrifugal force is balanced by the inward gravitational force.
In the case of the highly magnetized stars $R_{\mathrm K}$ is lower than the Alfv\'{e}n radius,
leading to the formation of a centrifugal magnetosphere \citep{petit_etal13}.
For HR\,5907 $R_{\mathrm K} \approx 1.5$ R$_{\ast}$ versus $<R_{\mathrm A}> \approx 16.5$ R$_{\ast}$ (Table~\ref{deriv_par}),
that is compatible with the building of a CM.
The accumulation of cold plasma in the CM}  
could be a source of anisotropic thermal free-free absorption, 
that further affects the non-thermal radio continuum emission.  
{The CM should have essentially no material between the star and $R_{\mathrm K}$ \citep{petit_etal13},
hence this can be easily modeled like a ring with inner radius equal to $R_{\mathrm K}$.}

To explain the observed behavior of the HR\,5907 radio emission,
we analyze the effect 
{that an equatorial ring}
of cold and dense thermal plasma
could have on the radio emission from the stellar magnetosphere.  
{The scale height of this cold ring decreases asymptotically outward.
In the case of an almost aligned dipole (like HR\,5907)
the asymptotic scale height can be derived by using the relation
$h_{\infty}=R_{\mathrm K} \sqrt{(2 k_{\mathrm B} T_{\mathrm {ring}} / \mu) / (3 G M_{\ast} / R_{\mathrm K})}$ \citep{townsend_owocky05}, 
where $k_{\mathrm B}$ is the Boltzmann constant 
and $\mu$ is the mean molecular weight (assumed equal to half proton mass for a fully pure hydrogen gas).
The temperature of the plasma ring has been fixed to $T_{\mathrm {ring}}=10^4$ K,
according to the temperature estimation of the cold plasma 
located inside the magnetosphere of HR\,5907 \citep{grunhut_etal12a}.
In the case of HR\,5907 $h_{\infty} \approx 0.03$ R$_{\ast}$.
}

In the synthetic model, that describes a typical magnetosphere of the early-type magnetic stars 
{(pictured in Fig.~\ref{modello}) an equatorial ring of cold plasma, assumed with a constant thickness, has been added.
The thickness of this cold ring has been conservatively assumed equal to 0.3 R$_{\ast}$, 
that is ten times the asymptotic scale height above estimated.
The spatial resolution of the finest grid spacing (0.1 R$_{\ast}$, see Sec.~\ref{num_sim})
is then able to sample the assumed ring thickness with at least 3 grid points.}
Following \citet{grunhut_etal12a}, who studied the rotational modulation
of the H${\alpha}$ features measured in the HR\,5907 spectra 
(due to the presence of this cold circumstellar material), 
{the outer radius of the ring
has been chosen equal to 4.4 R$_{\ast}$,
whereas the inner radius has ben fixed equal to the Kepler coronation radius.
The adopted} number density was set at $10^{13.5}$ cm$^{-3}$.  
{The adopted density and the outer radius are the maximal} values given by
\citet{grunhut_etal12a}, for which the analysis assumes
a temperature of $10^4$ K.

\begin{figure}
\resizebox{\hsize}{!}{\includegraphics{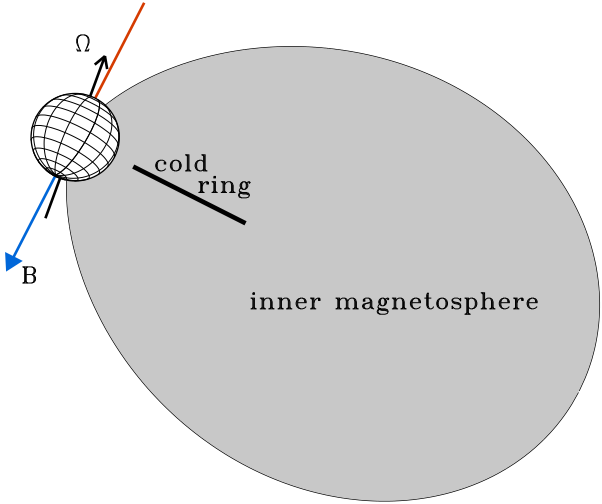}}
\caption{ 
Cartoon showing the magnetosphere of HR\,5907 where 
{the meridional cross section of a thin and dense equatorial cold ring has been pictured.}
The inner magnetosphere is indicated by the light-gray region.  
The angles used to picture the geometry of HR\,5907, inclination of the rotation axis $\Omega$
and obliquity of the magnetic axis, coincide with the values listed in Table~\ref{par_star}.
The magnetic field polarity is in accord with the magnetic axis orientation. 
The north pole is related to the hemisphere where the magnetic axis is outgoing from the star.}
\label{fig_cold_torus}
\end{figure}

The linear extension of the {thin dense equatorial ring} can be compared with the
size of the inner-magnetosphere by looking at Fig.~\ref{fig_cold_torus}.
The rotation axis inclination and the magnetic axis obliquity
displayed in the figure are those assumed for HR\,5907, 
listed in Table~\ref{par_star}.

The simulations performed {adding also a thin and dense cold ring}
within the deep magnetospheric layers of HR\,5907 are indistinguishable from
the simulations performed without it.
We conclude that, for the assumed magnetospheric geometry of HR\,5907, 
{an equatorial ring of dense and cold plasma 
(with an inner radius equal to 1.5\,R$_{\ast}$ and an outer radius equal to 4.4\,R$_{\ast}$)}
does not yield a significant effect on the simulated multiwavelength
radio light curves, neither for the total intensity (Stokes~$I$) nor
for the circularly polarized emission (Stokes~$V$).

\subsection{Auroral radio emission from HR\,5907?}
\label{radio_aurora}

The scenario that is able to simultaneously explain the gyro-synchrotron
radio emission and the non-thermal X-ray emission
{from the early type magnetic stars
(summarized in Appendix~\ref{appendix_mag_hot_star})} 
predicts that along with the X-ray aurora,
auroral radio emission should also be present. The auroral radiation
is a coherent radio emission powered by the electron cyclotron maser
(ECM) mechanism, that amplifies the radiation in the directions
almost perpendicular to the local magnetic field vector direction.
The ECM is driven by some type of unstable electron energy distribution,
in particular, the loss-cone \citep{wu_lee79,melrose_dulk82} or the
horseshoe \citep{winglee_pritchett86} distributions, that can be
developed by the fast electrons moving along the converging field
lines of a density-depleted magnetospheric cavity. In the case of
hot magnetic stars, the cavity coincides with the ``middle-magnetosphere''
{(see Fig.~\ref{modello}).}  
This coherent emission mechanism mainly
amplifies the extraordinary magneto-ionic mode giving rise to fully
polarized emission, with the polarization sense depending on the
magnetic field vector orientation
($RCP$ for emission from the northern side of the magnetosphere,
$LCP$ for emission from the southern side).

In the framework of the tangent plane beaming model, the coherent
auroral radio emission is mainly amplified tangentially to the cavity wall where it forms
\citep{louarn_lequeau96a,louarn_lequeau96b}.  
This is the case of the Earth's Auroral Kilometric Radiation (AKR)
\citep*{mutel_etal08,menietti_etal11}. The resulting auroral radio
emission will be constrained within a narrow beam pattern giving
rise to fully polarized broadband pulses, like a radio light-house.

\citet{leto_etal16} developed a 3D-model of the
stellar auroral radio emission that was able to successfully reproduce
the observed features of the CU\,Vir auroral pulses.  
This model has also been used to simulate the multi-wavelength auroral pulses
of the Ultra Cool Dwarf ($Sp=$M8.5) TVLM\,513-46546 \citep{hallinan_etal07}, allowing us to
constrain the overall topology of its magnetosphere \citep{leto_etal17b}.

The auroral radio emission at a fixed frequency arises from rings
located above the magnetic poles, at a height that depends on the
local value of the gyro-frequency $\nu_{\mathrm B}$, which is a
function of the magnetic field strength: $\nu_{\mathrm B} = 2.8
\times 10^{-3} B/{\mathrm G}$ GHz.  
In the case of a magnetosphere shaped by a simple dipole
the phase occurrence of the auroral pulses falls at phases almost
coinciding with the nulls of the effective magnetic field curve.  
This was observed in the case of the two already known 
hot magnetic stars characterized by auroral radio emission:
CU\,Vir ($Sp=$A0Vp) \citep{trigilio_etal00,trigilio_etal08,trigilio_etal11,ravi_etal10,lo_etal12};
and HD\,133880 ($Sp=$B8IVp) \citep*{chandra_etal15,barnali_etal18}.
In the case of HR\,5907 its effective magnetic field curve is always
negative \citep{grunhut_etal12a}, but the ORM geometry of this star
predicts that, at certain stellar rotational phases, the magnetic
field axis is almost perpendicular to the line of sight.  In principle
the auroral radio emission of HR\,5907 should be detectable.

\begin{figure}
\resizebox{\hsize}{!}{\includegraphics{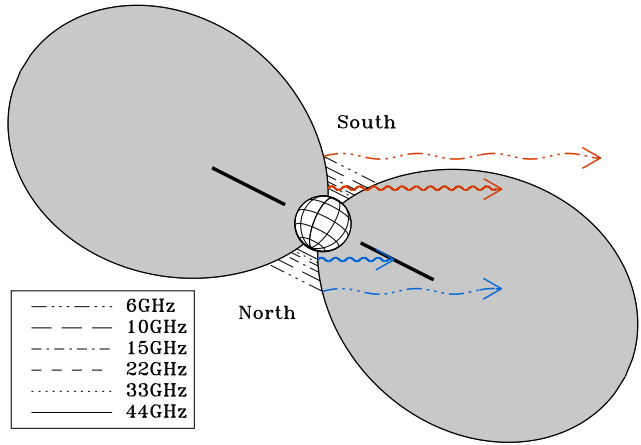}}
\caption{ 
Ray path of the auroral radio emission within the stellar magnetosphere
for HR\,5907. Note that the ray path of the auroral contribution arising from the northern hemisphere
traverses a larger amount of refracting/absorbing material {trapped within the inner magnetosphere.
Furthermore, some northern auroral radio frequencies could be blocked by the equatorial ring of cold plasma}.}
\label{cartoon_ecme}
\end{figure}

To test if the peculiar multi-wavelength radio light curves of
HR\,5907 are affected by the contribution of the stellar auroral
radio emission, we modeled the shape of the synthetic auroral radio
light curves arising from the magnetosphere of HR\,5907.  The
magnetospheric size was fixed at $R_{\mathrm A}=17$ R$_{\ast}$,
according to the average of the Alfv\'{e}n radii listed in
Table~\ref{deriv_par}.  The auroral rings within the HR\,5907
magnetosphere, tuned at the observed VLA radio frequencies,
are pictured in  Fig.~\ref{cartoon_ecme}.  The parameters that
control the auroral beam pattern are the ray path deflection and
the beaming angle, the latter centered on the plane tangent to the
auroral ring and containing the local magnetic field vector 
(see \citealp{leto_etal16} for details).  
These two angles are assumed as free parameters.  
We performed model simulations increasing the
auroral beam size from the case of emission constrained within a
narrow beam (auroral beam size equal to $20^{\circ}$) up to auroral
radio emission radiated isotropically.  The deflection angle of the
auroral ray path, with respect to the direction of the local magnetic
field vector, was assumed variable in the range $0^{\circ}$--$45^{\circ}$.

As a result of our simulations, we highlight that this elusive
coherent phenomenon could be detectable also for the HR\,5907 geometry.  
To be detectable, the auroral radio emission from HR\,5907
has to be radiated within a large beaming pattern 
(auroral beam size larger than $45^{\circ}$).  
The simulated fraction of the circularly polarized emission is comparable with the observed ones,
$<10$\% (see bottom panel of Fig.~\ref{fig_datav}), 
but this is possible only in the case of an absolutely symmetric auroral emission from the two opposite hemispheres.  
This is realized if the auroral radio emission
is radiated in directions exactly orthogonal to the local magnetic field vector orientation, 
both from the northern and southern side of the HR\,5907 magnetosphere.  
This ideal condition takes place if the stellar magnetic field topology is a pure dipole, 
to satisfy the required symmetry between the two hemispheres of opposite magnetic orientation, 
and if the auroral radio emission does not suffer any ray path deflection.

\begin{figure*}
\resizebox{\hsize}{!}{\includegraphics{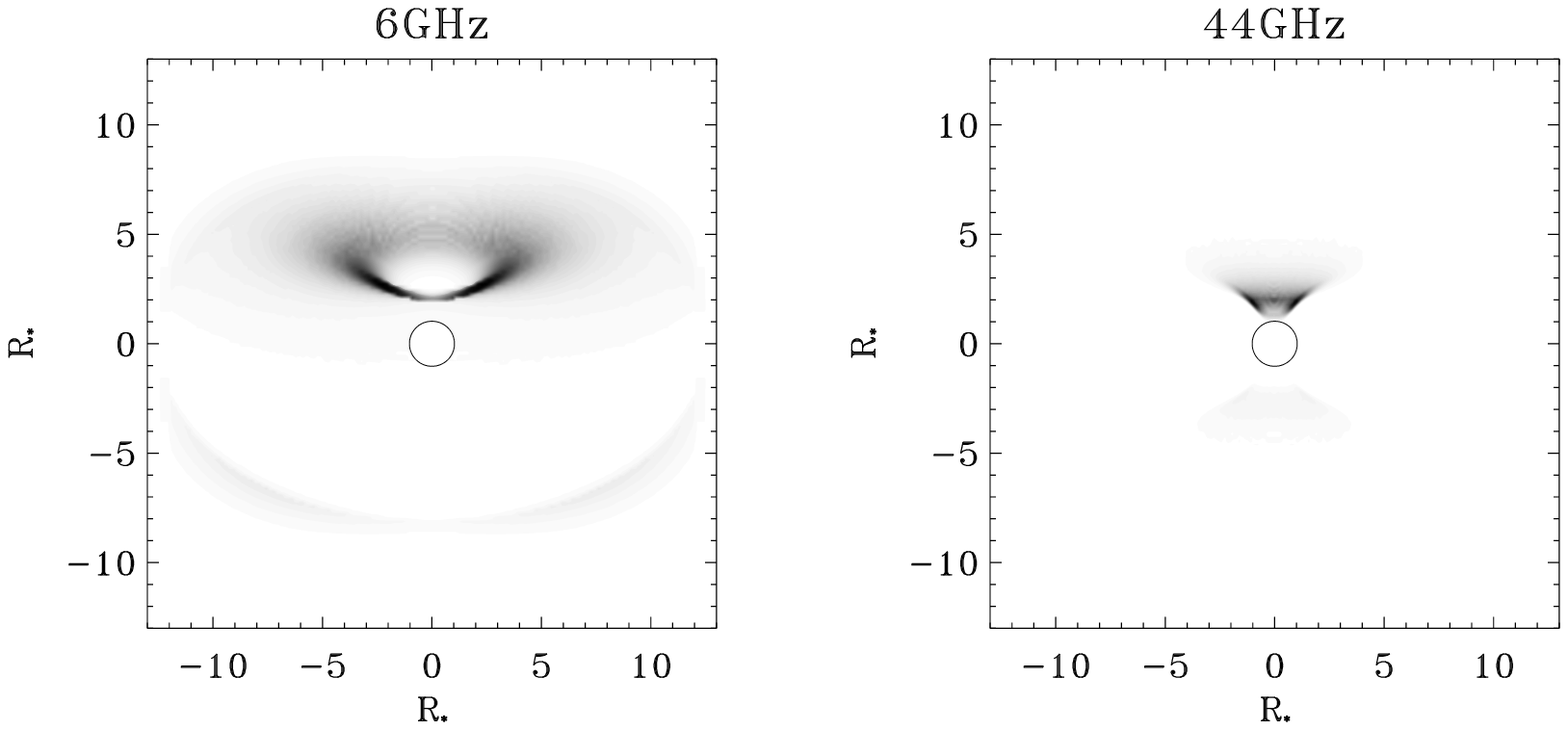}}
\resizebox{\hsize}{!}{\includegraphics{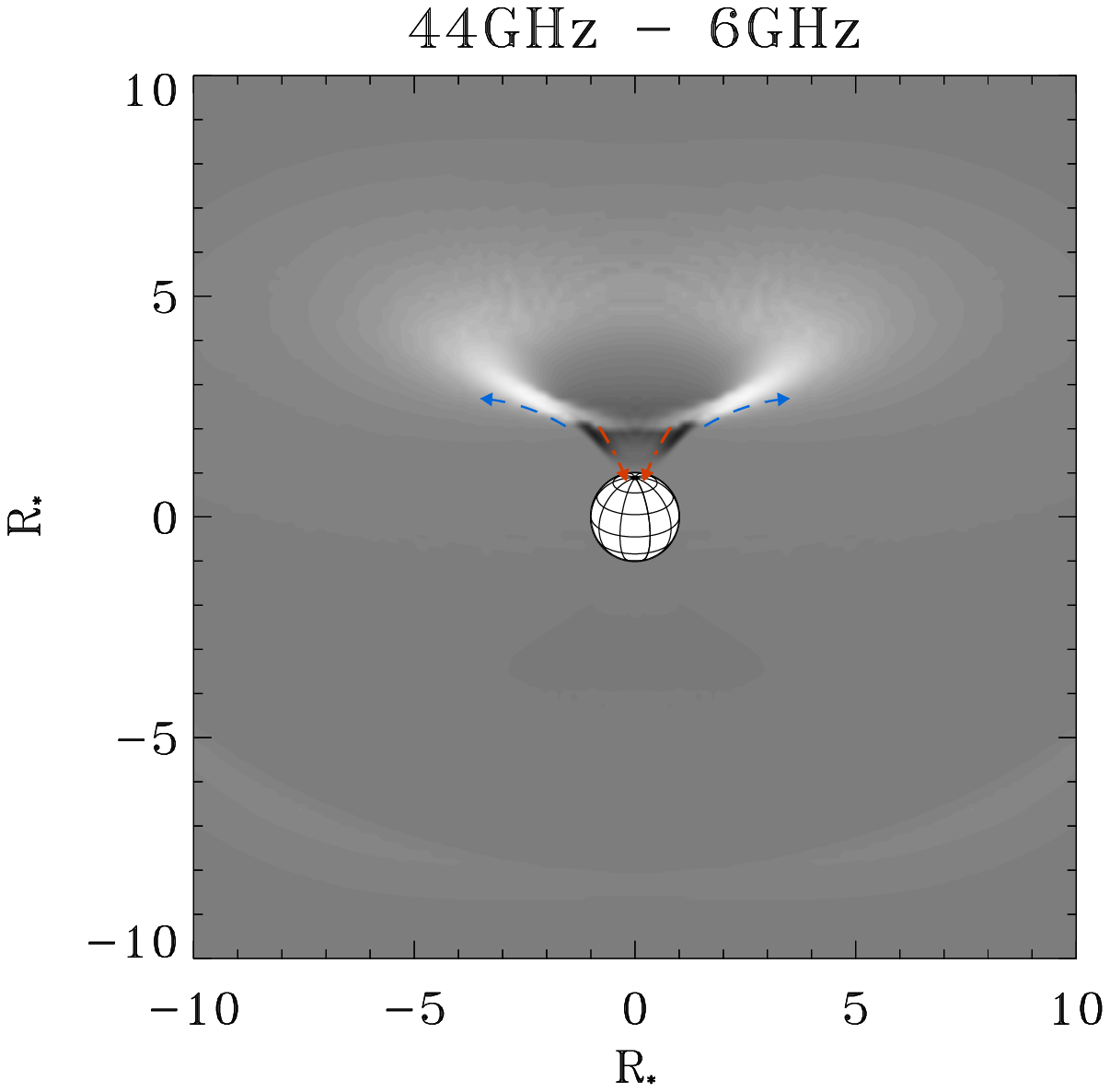}}
\caption{Top panels: HR\,5907 simulated maps at 6 (left panel) and 44 GHz (right panel), 
showing that radio emission forms closer to the star for shorter wavelengths.
Bottom panel: map difference between the 44 and the 6 GHz simulated maps. 
The magnetospheric region radiating at the two analyzed frequencies do not overlap.
The white area far from the star is related to the magnetospheric regions that mainly radiate at 6 GHz. 
The 44 GHz emission arises from
the black regions close to the star. 
The magnetic field topology is roughly sketched by the arrows pictured in the top panel.
The magnetic field vectors are mainly oriented toward the surface
in the regions close to the star, whereas they are outward oriented in the magnetospheric region
far from the star.
The displayed magnetic vector orientations are in accord with
the measured signs of the multi-wavelength circularly polarized radio emission.
}
\label{map6_44}
\end{figure*}

As already discussed \citep{leto_etal16}, the auroral radio emission
gives rise to features that are strongly circularly polarized if the ECM ray
path is not perfectly perpendicular to the local magnetic field
vector orientation. This could be a consequence of the intrinsic
nature of the elementary coherent emission mechanism, like the 
loss-cone driven ECM emission, that amplifies the radiation within a
hollow cone with an aperture that is a function of the electron
energy \citep{wu_lee79,melrose_dulk82}, or due to the presence of
thermal plasma trapped inside the stellar magnetosphere that refracts
the ECM rays traveling through \citep{trigilio_etal11}, or both.

{On the basis of the accepted scenario that describes the magnetosphere of the early-type magnetic stars, 
the inner-magnetosphere of such kind of stars is expected to be filled by thermal plasma.
It is then  plausible to expect that
the plasma trapped within the inner magnetosphere of HR\,5907 
has refracting/absorbing effects on the auroral radiation traveling through it.
In the case of HR\,5907 further frequency dependent blocking effects could be  
due to the presence of the equatorial ring of cold and dense plasma
that accumulates within the deep regions of its magnetosphere.
As previously discussed (Sec.~\ref{cold_torus}) 
such cold plasma has negligible effects to the simulated incoherent stellar radio emission,
but in the case of the coherent auroral radio emission its effects are not negligible.
The auroral radio emission at a given radio frequency, if present, 
originates from well defined auroral ring above the magnetic poles. 
Looking at Fig.~\ref{cartoon_ecme} it is clear that the}
auroral radio emission from HR\,5907 is not likely to be symmetric 
about the magnetic equator owing to path length and differential refractive effects.
For the geometry of HR\,5907,
the ray path length within the magnetospheric plasma is longer for
the auroral radiation arising from the northern hemisphere, as 
made clear in Fig.~\ref{cartoon_ecme}. This could give
rise to {refractive/absorbing} effects for the $RCP$ component of the HR\,5907 ECM
emission, with consequent lost of symmetry with the $LCP$ component,
that arises from the southern hemisphere.  

In the case of HR\,5907 there are also hints that its magnetic field topology is not a
simple dipole \citep{grunhut_etal12a}.  
{This gives rise} to asymmetric physical conditions
between the two stellar hemispheres of opposite magnetic field orientation,
that could significantly affect the elementary amplification mechanism
responsible of the stellar auroral radio emission.
{As an example, in the case of CU\,Vir, the prototype of a star} radiating auroral radio emission,
the existence of a non-dipolar magnetic field topology \citep{kochukhov_etal14}
has been ascribed to explain the detection of ECM emission from the northern hemisphere only
\citep{trigilio_etal00,trigilio_etal11}.

\subsection{The effects of a non-dipolar magnetic field on HR\,5907's radio emission}
\label{hr5907_mag}

The gyro-synchrotron emission mechanism is strongly sensitive to
the magnetic field strength, 
which varies with the distance from the star.  
Consequently, the radiation at a specific radio frequency band 
will be emitted mainly in a well localized region of the magnetosphere.  
The high-frequency radio emission originates close
to the stellar surface, where the field strength is strong, 
while the lower frequencies probe regions farther out.  
To highlight the different spatial location of the magnetospheric region where
the radio emission at a fixed frequency band originates, 
{we generated} synthetic radio maps of HR\,5907 at the two extremes
of the radio frequency range measured by the VLA, 
respectively 6 and 44 GHz.  
The HR\,5907 orientation corresponds to the 
extremum of the effective magnetic field curve.
The simulated maps at these two frequencies
are pictured in the top panels of Fig.~\ref{map6_44}.  
The synthetic maps {were created} 
using our 3D model assuming a dipolar field topology. 
The parameters that define the stellar magnetic field are
listed in Table~\ref{par_star}; the adopted model parameters are
listed in Table~\ref{deriv_par}.

{As is clear from} Fig.~\ref{map6_44}, the 44~GHz emission originates
close to the star, whereas the 6~GHz radio emission forms over a
larger, more extended volume.  This confirms how the gyro-synchrotron
radio emission is able to probe the physical conditions of the
stellar magnetosphere at different depths.  
To further highlight the different spatial 
{locations} 
of the regions where the 6 and 44
GHz radio emission originates, the 
{difference map} between the two
synthetic maps is also shown in the bottom panel of Fig.~\ref{map6_44}.
The 6 GHz emission arises from the large conical shaped region far
from the star ($>2$ R$_{\ast}$ above the stellar surface), 
{pictured in white,  whereas} the magnetospheric regions that mainly contributes
to the 44 GHz radio emission are the two filamentary black regions
close to the star, at about 1 R$_{\ast}$ above the stellar surface.
The high-frequency radio emission is then strongly sensitive to the
magnetospheric conditions close to the stellar surface,
like the non-thermal electron density, the magnetic field strength,
and its topology.

The radial dependence of the multi-polar magnetic field components are well known.  
In particular the magnetic field strength of a quadrupole decrease outward more steeply 
than a dipole, with $B_{\mathrm {quad.}} \propto r^{-5}$ 
versus $B_{\mathrm{dip.}} \propto r^{-3}$.  
On the basis of the above relations,
{a quadrupole} magnetic field mainly affects the field strength close to the stellar surface, 
whereas the layers far from the stellar surface are dominated by the simple dipole.
The ``hybrid" magnetic field topology is the vector sum of the dipole plus the quadrupole field.
The spatial dependence of the hybrid magnetic field strength is then a function of 
the strength at the stellar surface of the two field components (dipole plus quadrupole), 
{but is also a function of the orientation of the mutual magnetic moments. }
 
In the case of a hybrid field topology, the radial dependence of the magnetic field strength 
could significantly differ from the case of a simple dipole.
For a non-dipolar magnetic field topology,
it is then plausible to expect that the non-thermal electrons 
moving toward the stellar surface could meet magnetospheric layers 
where the local magnetic field strength is higher than the value
{expected in the case} of a simple dipole.
As a consequence, the magnetospheric regions
where the high order components of the magnetic field cannot be neglected 
will be a brighter source of incoherent gyro-synchrotron emission. 
This could qualitatively explain why the HR\,5907 total
intensity at 6 GHz is well reproduced by the simple dipolar model,
whereas the highest radio frequencies are not reproduced 
(see top panels of Fig.~\ref{simul_dati_ori}).

To determine whether the presence of a multipolar magnetic field component can explain 
the HR\,5907 flux level at $\nu \geq 10$ GHz 
(higher than $\approx 50\%$ compared to the simulated ones),
we performed new model simulations of the HR\,5907 gyro-synchrotron emission
{increasing by 50\%} the strength of the magnetic dipole moment.
Our model is not able to account 
{for a complex magnetic} field topology,
but we followed this procedure to simulate the effect 
of the increased magnetic field strength in the magnetospheric layers
where the radio emission at $\nu \geq 10$ GHz originates.
As a justification for a stronger magnetic field 
{than the estimated one,} 
we quote some papers that discuss the validity of the assumption that Stokes~$V$ profiles are equal 
to the first derivative of Stokes~$I$ profiles 
{for fast rotators \citep{scalia_etal17,leone_etal17}, an hypothesis }
that is at the basis of the measurements of the longitudinal component of magnetic fields. 
Moreover, there 
{are indication that chemical elements are not homogeneously distributed on the stellar surface \citep{grunhut_etal12a}, 
a condition that gives rise to a series}
of considerations regarding the recovered magnetic field strength
\citep*{stift_etal12,stift_leone17a,stift_leone17b}.
{At any rate, determination of the true magnetic field configuration of HR\,5907 is beyond the scope of this paper. These}
new simulations are only an exercise to help us 
understand how the magnetic field influences the incoherent
gyro-synchrotron emission from the stellar magnetosphere.

\begin{figure}
\resizebox{\hsize}{!}{\includegraphics{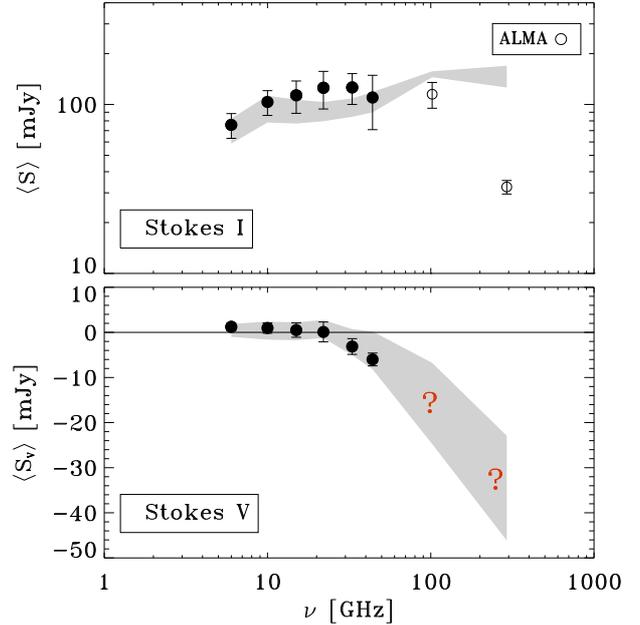}}
\caption{Results of the simulations performed increasing the magnetic field strength
and changing the dipole axis orientation at $\nu \leq 22$ GHz.
Top panel;
comparison between the observed and the simulated average spectrum of HR\,5907
for the total intensity (Stokes~$I$).
Bottom panel; comparison between the observed and the simulated average spectrum
for the circularly polarized emission (Stokes~$V$).
The measurements at the same frequency was averaged and represented by the filled dots.
This data has been already displayed in Fig.~\ref{sigi_sigv}.
The gray areas locate the range of the simulated values, constrained to the rotational phases covered by the observations.}
\label{sigi_sigv_sim}
\end{figure}

The new model simulations are compared with the observed average
spectrum of HR\,5907, Fig.~\ref{sigi_sigv_sim} top panel.  
{As a result of this exercise, we confirm that
increasing the magnetic field strength enhances the radio emission.}
The simulations performed using the amplified magnetic field strength highlights 
that the gyro-synchrotron emission model is able to well reproduce 
{the measured fluxes levels,} 
at least up to 102 GHz (see top panel of Fig.~\ref{sigi_sigv_sim}).
In fact, the simulated flux level at $\nu \geq 10$ GHz is closer to the observed ones.

The shapes of the simulated light curves remain
basically unchanged (flat light curves), like those pictured in the top panels of Fig.~\ref{simul_dati_ori}.
This is a further hint that the magnetosphere of HR\,5907 
strongly deviates from the pure-dipolar topology in the layers close to the stellar surface.
The large rotational modulation of the flux density observed at the high frequencies is
a clear indication that, as the star rotates, 
the magnetosphere of 
{HR\,5907 unveils} 
some anisotropic magnetic
structures characterized by local enhancement of the magnetic field strength.

Even if the rotational modulation of the HR\,5907 gyro-synchrotron
emission at 102 and 292 GHz is basically unknown, 
the average fluxes measured at the two ALMA bands
are displayed in the top panel of Fig.~\ref{sigi_sigv_sim}.
The average simulated flux density at 102 GHz is about 150 mJy, 
higher but reasonably close to the average of the measured fluxes 
listed in Table~\ref{alma_data_hr5907} (115 mJy).
The differences between the simulations and the observations become dramatic at 292 GHz.
The average simulated emission at 292 GHz is about $150$ mJy, versus an observed average flux of 32.5 mJy.  

\begin{figure*}
\resizebox{\hsize}{!}{\includegraphics{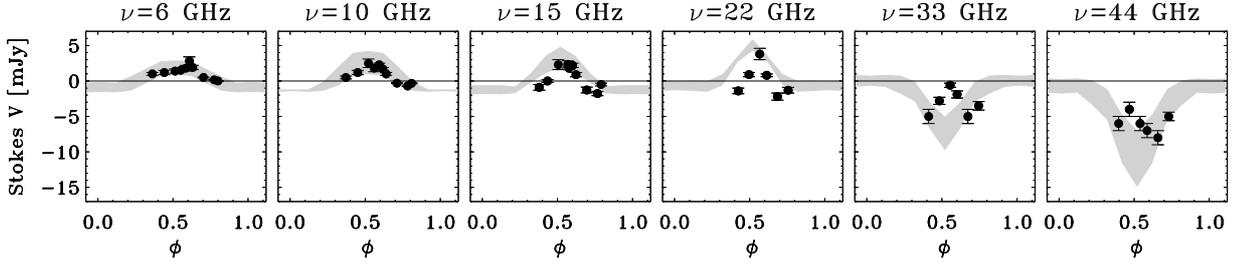}}
\caption{Light curves of the circularly polarized emission (Stokes~$V$) compared with the simulated ones.
The shaded areas are the envelopes of the simulated light curves.
The synthetic light curves 
{displayed in the figure} 
try to simulate the case of a non-dipolar magnetic field topology.
The strength of the magnetic moment has been enhanced of 50\% for the simulations at $\nu \geq 10$ GHz.
The stellar geometry parameters assumed for the simulations at 33 and 44 GHz are listed in Table~\ref{par_star}.
The simulations at $\nu \leq 22$ GHz were instead performed assuming 
the opposite dipole orientation, exchange of the north with the south pole.}
\label{simul_dati_v}
\end{figure*}

The flux drop measured at 292 GHz 
could be related to the true density radial profile of the non-thermal
electrons within the magnetospheric flux tubes where they freely propagate.  
In fact, the radiative and collisional loss effects
could significantly affect the density of the non-thermal electrons
very close to the stellar surface, where the emission at 292 GHz
is expected to arise. 

The puzzling nature of the HR\,5907 radio emission is further
complicated by the behavior of its circularly polarized emission,
quantified by the Stokes~$V$ parameter.  
As already discussed, the Stokes~$V$ sign is correlated with the sign of the effective magnetic field,
{as an example} positive Stokes~$V$ is produced in regions with the magnetic field vector 
mainly oriented toward the observer ($B_{\mathrm e} > 0$).
The measured HR\,5907 effective magnetic field is always negative \citep{grunhut_etal12a}.
The magnetic field vectors are then mainly oriented 
{toward the star, thus the} 
expected signs of the Stokes~$V$ measurements will be negative.  
As already discussed in Sec.~\ref{circ_pol},
the HR\,5907 radio emission at $\nu\leq 22$ GHz
is anti-correlated with the sign of the effective magnetic field.

To try to explain the anomalous behavior of the HR\,5907 circularly polarized radio emission, 
{we have arbitrarily changed}
the magnetic field topology as a function of the height above the stellar surface.
The regions where the radio emission in the range 6--22 GHz originates are 
{assumed characterized by 
magnetic field vectors mainly oriented toward the observer, whereas
the regions close to the star maintain the original orientations of the magnetic field vectors.}
The arrows
{superimposed on the difference map} 
(spatial brightness distribution at 44 GHz minus brightness  at 6 GHz, bottom panel of Fig.~\ref{map6_44}) 
are used to visualize 
{such a complicated} magnetic field topology. 
The vectors pictured by the dot dashed line (in red in the online color version of the Fig.~\ref{map6_44})  
are oriented toward the stellar surface, and are in accordance
with the effective magnetic field measurements (negative $B_{\mathrm e}$),
the vectors pictured using the dashed line have instead opposite 
{orientations} (in blue in the online color version of the Fig.~\ref{map6_44}).

New model simulations at $\nu=6$, 10, 15 and 22 GHz
{have now been performed} flipping the dipole axis orientation.
The simulated total intensity and magnitude of the circularly polarized emission remain unchanged,
whereas the sign of the simulated Stokes~$V$ reverses.
{In this way,} 
the overall behavior of the rotational modulation of the circularly polarized emission 
of HR\,5907 is {better} reproduced by the simulations, see Fig.~\ref{simul_dati_v}.
The agreement between observations and simulations is also
highlighted by comparing the average spectral dependence of 
the measured Stokes V emission with the observed ones, see bottom panel of Fig.~\ref{sigi_sigv_sim}.
{We want to remember that the assumed magnetic filed topology is purely empirical.
This exercise was only used to obtain indirect indications regarding  
the true magnetic field topology of HR 5907,
that is surely much more complex than a simple dipole.
In fact, the model reproduces the overall Stokes~$V$ reversal as a function of the frequency,
but is not able to reproduce the detailed behavior observed at 33 and 44 GHz 
(see the two right panels of Fig.~\ref{simul_dati_v}).
In particular, a simple dipolar topology predicts 
the maximum fraction of the circularly polarized emission coinciding with the rotational phase 
where the maximum effective magnetic field reaches its maximum strength ($\phi \approx 0.5$), 
this is not evidenced by the 33 and 44 GHz measurements.}

The model predictions at the ALMA bands have been also performed.
These simulations extend the expected spectrum of the circularly polarized emission
from HR\,5907 close to the sub-millimeter wavelength range, Fig.~\ref{sigi_sigv_sim} bottom panel.
The simulated circularly polarized emission at the two ALMA frequencies
is about 10\% at 102 GHz and 20\% at 292 GHz.
But no comparison with the observations can be done due to the missing
information of the measured fraction of the circularly polarized emission.

\section{Summary and Conclusions}
\label{sum}

The radio emission of the early B-type magnetic star HR\,5907
measured at many frequencies, from 6 up to 292 GHz, combined with X-ray spectroscopy
provides us with the opportunity to probe the physical conditions of its magnetosphere.

The fluxes at the frequencies from 6 to 44 GHz were measured by the VLA interferometer.
These radio observations cover a large fraction of the stellar rotational phases,
showing that the gyro-synchrotron radio emission of HR\,5907 is strongly variable as a consequence of the stellar rotation,
in accordance with a RRM shaped like an Oblique Rotator.
This is the accepted scenario that well explains the rotational modulation of the radio emission 
observed in many hot magnetic stars. 

In this paper we report the first  measurements of HR\,5907 with the millimeter/submillimeter ALMA interferometer. 
ALMA clearly detected HR\,5907 at 102 and 292 GHz,
allowing to constrain the source spectrum up to the boundary of the sub-millimeter wavelength range.
The gyro-synchrotron emission mechanism is strongly sensitive to the magnetic field strength and orientation.
Consequently, the radiation within a specific radio frequency band is emitted mainly in a well-localized layer of the magnetosphere. 
Higher frequency emissions are generated close to the stellar surface, where the field strength is higher, 
whereas lower frequencies probe regions farther out.

The behavior of the gyro-synchrotron emission at the millimeter 
and sub-millimeter spectral range is poorly studied for 
{this kind of star.}
The detection of HR\,5907 at the millimeter wavelengths indicates
that in this star the population of non-thermal electrons, 
that radiate for gyro-synchrotron emission mechanism,
efficiently penetrates to the magnetospheric layers very close to the stellar surface.

The radio/millimeter observations of HR\,5907 were complemented by the X-ray spectroscopy. 
The {\em Chandra} ACIS-S spectrum of HR\,5907 was well 
{fitted by a spectral model that combined a thermal and non-thermal components.}
Following the MCWS model, the thermal component originates from plasma heated up to 10\,MK.
On the other hand, the non thermal emission 
component is best  explained as auroral X-ray emission induced  by the 
non-thermal electrons that impact with the stellar surface close to the polar caps.

We have previously realized a 3D model able to simulate the gyro-synchrotron radiation
arising from the RRM of the hot magnetic stars.
The physical parameters derived from the X-ray spectrum fitting
have been used to constrain the model free parameters needed to simulate
the non-thermal radio emission of HR\,5907.
However, we were not able to successfully reproduce all the features observed 
at the radio/millimeter wavelengths.

To explain the disagreement between observations and simulation we considered the presence 
of an equatorial {ring of dense and cold plasma}
within the magnetosphere of HR\,5907.
We concluded that for the geometry of HR\,5907 this 
{ring has no effect on} the radio emission propagation.
We also analyzed the case of a possible contribution to the radio emission of HR\,5907 of the coherent auroral radio emission.
This hypothesis has also been discarded.

Then we performed a semi-quantitative analysis 
on the effects of a magnetic field topology more complex than a simple dipole 
to the radio emission from the HR\,5907 magnetosphere.
This analysis highlights that the magnetic field
has a key role in explaining the observed features at the radio wavelengths.
Even if we are not able to quantify the true topology and strength of the HR\,5907 magnetic field,
we provide an indirect confirmation of the suggested existence of a multipolar magnetic field component \citep{grunhut_etal12a}. 

We point out that the radio emission from the hot magnetic stars studied so far are seen to be time stable, 
indicating that the physical mechanisms causing them are in steady state, 
making these magnetospheres ideal laboratories for the study of the plasma processes
responsible for the phenomena detected from the radio to the X-rays.
These phenomena have also been detected in the magnetospheres of totally different
objects, such as the magnetized planets of the solar system and the brown dwarfs.
From this point of view the measurements of the millimeter and sub-millimeter 
light curves of the hot magnetic stars provide strong constraints on the physical conditions
in the deep magnetospheric layers,
{where the auroral X-ray emission also originates.}

\section*{Acknowledgments}
We thank the referee for their very constructive comments that helped to improve the paper. 
LMO acknowledges support by the DLR grant 50\,OR\,1508.
The National Radio Astronomy Observatory is a facility of the National Science 
Foundation operated under cooperative agreement by Associated Universities, Inc.
This work has extensively used  the NASA's Astrophysics Data System, 
and the SIMBAD database, operated at CDS, Strasbourg, France. 
Part of the scientific results reported in this article are based on
data obtained from the {\em Chandra} Data Archive.
This paper makes use of the following ALMA data: ADS/JAO.ALMA\#2011.0.00001.CAL.
ALMA is a partnership of ESO (representing its member states), 
NSF (USA) and NINS (Japan), together with NRC (Canada), MOST and ASIAA (Taiwan), 
and KASI (Republic of Korea), in cooperation with the Republic of Chile. 
The Joint ALMA Observatory is operated by ESO, AUI/NRAO and NAOJ.

\appendix
\section{List of the radio measurements}
The radio measurements of HR\,5907 presented in this paper are listed in Table~\ref{radio_data_hr5907},
column 3 total intensity (Stokes~$I$) and column 4 circularly polarized intensity (Stokes~$V$), with the corresponding uncertainty.
The average HJD of each observing scan (column 1) and
the corresponding rotational phase (column 2) have been also reported.
The millimeter measurements performed with the ALMA interferometer are listed in Table~\ref{alma_data_hr5907}.
The phases of the periodic variability are computed according to the ephemeris given by Eq.~\ref{effemeridi}.

\begin{table} 
\caption{VLA logbook and radio measurements of HR\,5907} 
\label{radio_data_hr5907} 
\begin{tabular}{ccll} 
\hline 
HJD        &$\phi$      &~~~~~S     &~~~~~~S${_{V}}$             \\
2457000+   &            &~~(mJy)    &~~~(mJy)                                             \\
\hline                                                                                           
\multicolumn{4}{c}{$\nu=6$ GHz}                                                                       \\            
\hline                                                                                                 
90.8973  & 0.5510       & ~~88 \s{($\pm1$)}           & ~~~1.5 \s{($\pm0.2$)}     \\     
90.9382  & 0.6317       &  ~~74.6 \s{($\pm0.8$)}       & ~~~1.9 \s{($\pm0.3$)}      \\
90.9746  & 0.7032       &  ~~63.6 \s{($\pm0.6$)}       & ~~~0.50 \s{($\pm0.08$)}      \\
91.0110  & 0.7748       &  ~~56.6 \s{($\pm0.5$)}       & ~~~0.16 \s{($\pm0.06$)}      \\
91.0235  & 0.7993       &  ~~55.0 \s{($\pm0.4$)}       & ~~~0.00 \s{($\pm0.04$)}      \\
97.9175  & 0.3628       &  ~~80 \s{($\pm2$)}           & ~~~1.0 \s{($\pm0.2$)}      \\
97.9578  & 0.4421       &  ~~83.1 \s{($\pm0.7$)}       & ~~~1.2 \s{($\pm0.3$)}      \\ 
97.9941  & 0.5136       &  ~~86 \s{($\pm1$)}           & ~~~1.4 \s{($\pm0.3$)}      \\    
98.0305  & 0.5852       &  ~~86 \s{($\pm3$)}           & ~~~1.8 \s{($\pm0.4$)}      \\
98.0426  & 0.6090       &  ~~84 \s{($\pm1$)}           & ~~~2.8 \s{($\pm0.6$)}       \\
\hline                                                                                                
\multicolumn{4}{c}{$\nu=10$ GHz}                                                                      \\
\hline                                                                                                
90.9017  & 0.5598      &121 \s{($\pm2$)}           & ~~~1.8 \s{($\pm0.5$)}      \\ 
90.9423  & 0.6397      &101 \s{($\pm2$)}           & ~~~1.0 \s{($\pm0.2$)}      \\
90.9787  & 0.7113      &~~87 \s{($\pm2$)}            & $-0.30$ \s{($\pm0.07$)}      \\  
91.0151  & 0.7829      &~~77 \s{($\pm2$)}            & $-0.7$ \s{($\pm0.1$)}      \\
91.0279  & 0.8081      &~~78 \s{($\pm2$)}            & $-0.34$ \s{($\pm0.07$)}      \\
97.9216  & 0.3709      &104 \s{($\pm4$)}           & ~~~0.5 \s{($\pm0.2$)}      \\
97.9619  & 0.4502      &110 \s{($\pm2$)}           & ~~~1.2 \s{($\pm0.3$)}      \\
97.9982  & 0.5217      &119 \s{($\pm2$)}           & ~~~2.5 \s{($\pm0.6$)}      \\ 
98.0346  & 0.5932      &120 \s{($\pm3$)}           & ~~~2.3 \s{($\pm0.2$)}      \\
98.0467  & 0.6171      &117 \s{($\pm2$)}           & ~~~1.7 \s{($\pm0.2$)}      \\
\hline                                                                                                
\multicolumn{4}{c}{$\nu=15$ GHz}                                                                      \\
\hline                                                                                                
90.9060  & 0.5682      &133 \s{($\pm5$)}             & $-1.1$ \s{($\pm0.2$)} \\
90.9341  & 0.6236      &122 \s{($\pm4$)}             & $-0.10$ \s{($\pm0.05$)}      \\
90.9705  & 0.6952      & ~~95 \s{($\pm2$)}             & ~~~0.26 \s{($\pm0.04$)}      \\
91.0069  & 0.7667      & ~~79 \s{($\pm3$)}             & ~~~1.1 \s{($\pm0.1$)}        \\
91.0192  & 0.7909      & ~~72 \s{($\pm5$)}             & ~~~1.6 \s{($\pm0.3$)}        \\
97.9255  & 0.3786      &107 \s{($\pm5$)}             & ~~~2.3 \s{($\pm0.4$)}        \\
97.9527  & 0.4340      &114 \s{($\pm2$)}             & ~~~1.9 \s{($\pm0.4$)}        \\
97.9900  & 0.5056      &135 \s{($\pm2$)}             & ~~~0.9 \s{($\pm0.2$)}        \\
98.0264  & 0.5771      &143 \s{($\pm2$)}             & ~~~0.40 \s{($\pm0.04$)}      \\
98.0387  & 0.6013      &132 \s{($\pm6$)}             & $-0.7$ \s{($\pm0.2$)}        \\
\hline                                                                                                
\multicolumn{4}{c}{$\nu=22$ GHz}                                                                      \\
\hline                                                                                                
90.9290  & 0.6136      &149 \s{($\pm2$)}             & ~~~0.8 \s{($\pm0.2$)}      \\
90.9654  & 0.6851      &101 \s{($\pm9$)}             & $-2.2$ \s{($\pm0.5$)}      \\
91.0018  & 0.7567      & ~~78 \s{($\pm8$)}             & $-1.3$ \s{($\pm0.4$)}      \\
97.9486  & 0.4240      &120 \s{($\pm10$)}            & $-1.4$ \s{($\pm0.4$)}      \\
97.9849  & 0.4955      &141 \s{($\pm8$)}             & ~~~0.9 \s{($\pm0.3$)}      \\
98.0213  & 0.5670      &162 \s{($\pm7$)}             & ~~~3.8 \s{($\pm0.8$)}      \\
\hline                                                                                                
\multicolumn{4}{c}{$\nu=33$ GHz}                                                                      \\
\hline                                                                                                
90.9228  & 0.6013      &160 \s{($\pm20$)}            & $-1.9$ \s{($\pm0.5$)}      \\
90.9592  & 0.6729      &112 \s{($\pm8$)}             & $-5$ \s{($\pm1$)}          \\
90.9955  & 0.7444      & ~~97 \s{($\pm7$)}             & $-3.5$ \s{($\pm0.6$)}      \\
97.9423  & 0.4117      &103 \s{($\pm6$)}             & $-5$ \s{($\pm1$)}          \\
97.9787  & 0.4832      &130 \s{($\pm10$)}            & $-2.8$ \s{($\pm0.5$)}      \\
98.0150  & 0.5548      &160 \s{($\pm20$)}            & $-0.6$ \s{($\pm0.3$)}      \\
\hline                                                                                                
\multicolumn{4}{c}{$\nu=44$ GHz}                                                                      \\
\hline                                                                                                
90.9153  & 0.5865      &165 \s{($\pm3$)}             &$-7$ \s{($\pm1$)}      \\
90.9516  & 0.6580      &110 \s{($\pm8$)}             &$-8$ \s{($\pm1$)}      \\
90.9880  & 0.7296      &~~75 \s{($\pm10$)}             &$-5.0$ \s{($\pm0.6$)}      \\
97.9348  & 0.3969      &~~60 \s{($\pm9$)}              &$-6$ \s{($\pm1$)}      \\
97.9711  & 0.4684      &110 \s{($\pm5$)}             &$-4$ \s{($\pm1$)}      \\
98.0075  & 0.5400      &140 \s{($\pm7$)}             &$-6$ \s{($\pm1$)}      \\
\hline                                                                                                             

\end{tabular}     
\end{table} 

\begin{table} 
\caption{ALMA logbook and millimeter measurements of HR\,5907} 
\label{alma_data_hr5907} 
\begin{tabular}{ccl  ccl} 
\hline 
HJD        &$\phi$      &~~~~~S                &HJD        &$\phi$      &~~~~~S    \\
2457000+   &            &~~(mJy)              & 2457000+   &            &~~(mJy)     \\
\hline                                                                                           
\multicolumn{3}{c}{$\lambda=3$ mm ($\nu=102$ GHz)}          &\multicolumn{3}{c}{$\lambda=1$ mm ($\nu=292$ GHz)}              \\            
\hline                                                                                                 
237.3487  & 0.6847       & 128 \s{($\pm7$)}            &237.3491  & 0.6854      &32 \s{($\pm3$)}   \\     
237.3756  & 0.7377       & 102 \s{($\pm5$)}            &237.3752  & 0.7369      &33 \s{($\pm3$)}      \\
\hline                                                                                                
\end{tabular}     
\end{table} 

\section{Radio and X-ray emission from the early-type magnetic stars}
\label{appendix_mag_hot_star}

\begin{figure}
\resizebox{\hsize}{!}{\includegraphics{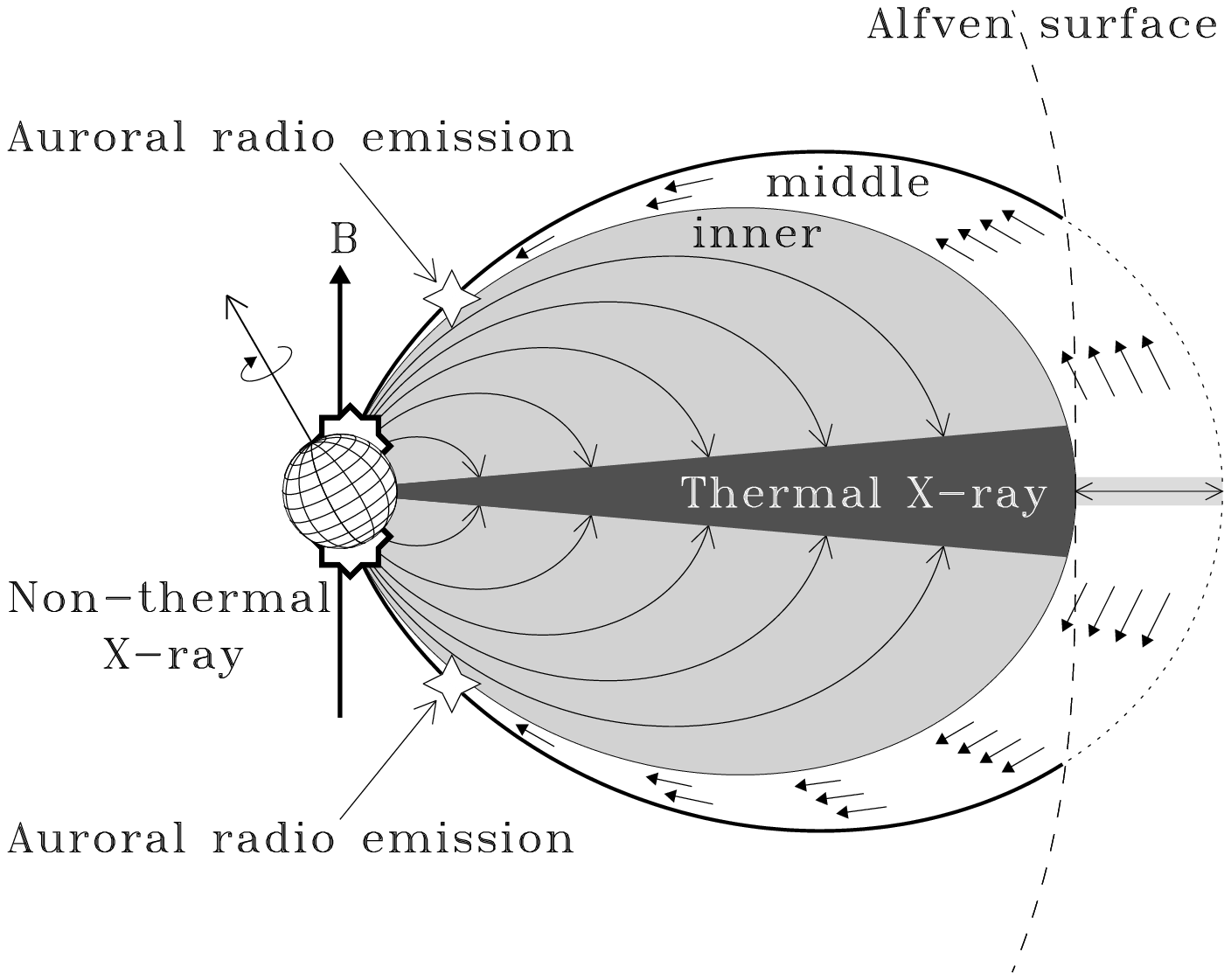}}
\caption{{Magnetosphere of a dipole shaped early-type magnetic star.
The amplitude of the misalignment of the dipole and rotation axis is arbitrary.
The small vectors in the middle magnetosphere oriented toward the star represent the streams of non-thermal electrons.
}}
\label{modello}
\end{figure}

The scenario able to explain the radio and the X-ray emission
from a dipolar shaped magnetosphere of a typical hot magnetic star is pictured in Fig.~\ref{modello}.
The wind plasma streams, arising from the two opposite hemispheres,
move along the dipolar magnetic field lines
and collide at the magnetic equator. This gives rise to a shock that heats the plasma,
making this hot plasma able to radiate thermal X-rays 
(pictured in Fig.~\ref{modello} by using the dark gray area close to the magnetic equatorial plane).
At the higher magnetic latitudes, the wind plasma accumulates remaining trapped 
within the inner magnetospheres (the light gray area in Fig.~\ref{modello}).
At the distance equal to the Alfv\'{e}n radius ($R_{\mathrm A}$) the magnetic field strength is not able
to confine this thermal plasma and the co-rotation breakdown takes place.
The equatorial regions outside the Alfv\'{e}n surface are site
of magnetic reconnection events, with consequents generation of fast electrons. 
These non-thermal electrons freely propagate toward the star
radiating at the radio wavelengths by incoherent gyro-synchrotron emission mechanism.
The dipolar cavity where the radio emission originates is named middle magnetosphere.
The fast electrons streams that collide with the stellar surface
radiate non-thermal X-rays by thick-target bremsstrahlung emission.
The non-thermal electron population outward reflected could then develop
the unstable loss-cone energy distribution.
These electrons are then responsible of the highly beamed stellar auroral radio emission 
due to the coherent electron cyclotron maser emission mechanism.

\end{document}